\documentclass[12pt,letterpaper]{article}
\usepackage{jheppub}
\usepackage[utf8]{inputenc}
\usepackage{amsthm}
\usepackage{dcolumn}
\usepackage{cancel}
\usepackage{booktabs}
\usepackage{multirow}
\usepackage{esvect}
\usepackage{ytableau}
\usepackage{stackengine}[2013-10-15]
\usepackage{amsmath}
\newcommand{\bartek}{}

\newcolumntype{C}[1]{>{\centering\let\newline\\\arraybackslash\hspace{0pt}}m{#1}}

\def\be{\begin{equation}}
\def\ee{\end{equation}}
\def\ba#1\ea{\begin{align}#1\end{align}}
\def\bg#1\eg{\begin{gather}#1\end{gather}}
\def\bm#1\em{\begin{multline}#1\end{multline}}
\def\bmd#1\emd{\begin{multlined}#1\end{multlined}}

\def\({\left(}
\def\){\right)}
\def\[{\left[}
\def\]{\right]}
\def\<{\langle}
\def\>{\rangle}

\mathchardef\arr="017E 
\newcommand\dvec[1]{\setbox0=\hbox{$#1$}\mbox{\hbox to 0pt{\hbox to \wd0{\hss\raisebox{-1.95ex}[0ex][0ex]{$\arr\;$}\hss}\hss}}\box0}

\begin{document}

\title{Boundary and bulk notions of transport \\ in the AdS$_3$/CFT$_2$ correspondence}
\author[a]{Bowen Chen}
\author[a]{Bart{\l}omiej Czech}
\affiliation[a]{Institute for Advanced Study, Tsinghua University, Beijing 100084, China}
\author[b]{Jan de Boer}
\affiliation[b]{Institute for Theoretical Physics and Delta Institute for Theoretical Physics, \\\mbox{University of Amsterdam}, PO Box 94485, 1090 GL Amsterdam, The Netherlands}
\author[c]{Lampros Lamprou}
\affiliation[c]{Department of Physics and Astronomy, \\ 
University of British Columbia, Vancouver, BC V6T 1Z1, Canada}
\author[a]{Zi-zhi Wang}
\emailAdd{chenbw95@gmail.com, bartlomiej.czech@gmail.com, j.deboer@uva.nl, llamprou90@gmail.com, zizhi.wang08@gmail.com}

\abstract{
We construct operators in holographic two-dimensional conformal field theory, which act locally in the code subspace as arbitrary bulk spacelike vector fields. Key to the construction is an interplay between parallel transport in the bulk spacetime and in kinematic space. We outline challenges, which arise when the same construction is extended to timelike vector fields. We also sketch several applications, including boundary formulations of the bulk Riemann tensor, dreibein, and spin connection, as well as an application to holographic complexity.
}

\maketitle
\thispagestyle{empty}
\setcounter{page}{1}

\section{Introduction}
The last decade has seen remarkable progress in understanding bulk reconstruction in holographic duality through the lens of quantum information theory; see \cite{rev1, rev2, rev3, rev4, rev5} for recent reviews. Several of the underlying works naturally belong in one category: they explain the emergence of Poincar{\'e} symmetry in the bulk by constructing bulk vector fields. An incomplete list of results, which reconstruct bulk vector fields as boundary operators using information theoretic tools, includes:
\begin{itemize}
\item the JLMS relation \cite{jlmsref},
\item modular parallel transport \cite{modberry,sewingkit} and modular scrambling modes \cite{modularchaos},
\item the worldline of a massive particle \cite{bulktime}.
\end{itemize}
This paper adds an entry to this list. We construct a field theory operator, which acts as a given spacelike vector field in a neighborhood of any one of its integral curves.

Our construction exploits an interplay between three notions of transport along a bulk spacelike curve. One of them will be the desired bulk vector field, another---modular parallel transport, and the third will be their sum. To aid the reader in following the logic, here we describe an analogy, which in recent years has become familiar to all drivers:

\paragraph{Analogy: GPS display}
Consider a GPS unit, which displays the road ahead in the default `Heading up' setting. This means that whenever the car makes a turn, the display rotates accordingly. The driver goes on a joy ride along trajectory $\mathcal{C}$, which is topologically a circle. We assume that the journey begins and ends at the same location on Earth, and at the same azimuth (orientation relative to North). It is clear that the GPS unit displays the same view before and after the journey.

The geometric transformation performed by the navigation device during the journey is not parallel transport along the driver's trajectory. The latter generates rotational holonomies---a fact known as the Coriolis effect. If not parallel transport, what is the transformation applied by the navigation unit?

A moment's thought reveals that the generator of the GPS transformation must be decomposable into two components: parallel transport along the curve plus an extra rotation about the car's instantaneous position. The amount of rotation is set to precisely counter the Coriolis effect induced by parallel transport. It should be clear that the right amount of rotation is fixed by the extrinsic curvature of the trajectory $\mathcal{C}$. After all, if the driver completed a great circle (a geodesic on the surface of the Earth), no extra rotation would be necessary. 

\paragraph{Strategy} The above example illustrates that the geometric problem we face concerns a set of three connected concepts. To set the notation, we will be looking for three operator-valued differential forms obeying:
\begin{equation}
A d\lambda = Z d\lambda + V d\lambda\,,
\label{ourproblem}
\end{equation} 
where $\lambda$ parameterizes a curve $\mathcal{C}$. The form $A d\lambda$ will generate the kind of flat transport, which is illustrated by the GPS example. Under suitable assumptions, the form $Z d\lambda$ will generate bulk parallel transport along a given curve. The form $V d\lambda$ will generate instantaneous rotations, which undo the `Coriolis effect.' 

Whereas $A d\lambda$, by construction, generates trivial holonomies, integrating $V d\lambda$ and $Z d\lambda$ gives rise to non-trivial holonomies. In the simplest settings we will find:
\begin{equation}
{\rm Pexp} \oint Z d\lambda = {\rm rotation}
\qquad {\rm and} \qquad
{\rm Pexp} \oint V d\lambda = {\rm translation}
\end{equation}
The second of these holonomies has been studied at length in the literature \cite{diffent, hholes, intgeo}. It is the modular Berry phase \cite{modberry,sewingkit}, which accumulates when we vary boundary regions (and their modular Hamiltonians) in an analogue of the Berry phase setup \cite{berryref}. The parameter space, which comprises modular Hamiltonians drawn from a common global state, is called kinematic space \cite{stereoscopic,jansversion}. The operator-valued differential form $V d\lambda$, which supplements bulk parallel transport $Z d\lambda$ to define the flat GPS-like transport, also generates parallel transport---but in kinematic space. The essence of our strategy is that the two distinct notions of parallel transport---in the bulk and in kinematic space---become flat when used in combination.

Determining forms $Z d\lambda$ given $V d\lambda$ has a number of practical applications. Finding a boundary operator that generates a bulk translation along a given curve is a first example. We also discuss several others: a link with the Chern-Simons formalism for AdS$_3$ gravity\footnote{Chern-Simons fields, which solve the equations of motion---that is, obey Einstein's equations---are flat $sl(2,\mathbb{R})$ connections. For this reason, they naturally correspond to the holonomy-free $A d\lambda$.} \cite{Witten:1988hc}, an avenue to computing the bulk Riemann tensor, spin connection and dreibein, and another application to holographic complexity. These applications are explained in Section~\ref{sec:appl}.

\paragraph{Organization}
Section~\ref{sec:rev} reviews the concept of modular parallel transport. Section~\ref{sec:maingen} sets up a new transport problem to be studied in this paper, which defines two operator-valued forms $Z d\lambda$ and $A d\lambda$. In Section~\ref{sec:main} we specialize to three bulk dimensions and find that the bulk action of $Z d\lambda$ (or an appropriate component of it; see Section~\ref{sec:notimerefl} for a detailed statement) is the same as bulk parallel transport along a given spacelike curve. We discuss explicit examples in pure AdS$_3$, though our result applies in general three-dimensional bulk geometries. Section~\ref{sec:discussion} contrasts our results with bulk parallel transport along timelike curves and discusses applications. 

\section{Review of modular parallel transport}
\label{sec:rev}

Consider two boundary regions $A = R(\lambda)$ and $B = R(\lambda + d\lambda)$ in a holographic CFT. The notation reflects an assumption that the two regions are related by a small shape variation, which is parameterized by a variable $\lambda$. The CFT is in some global state $\rho$, which may be pure or mixed, but which is dual to a definite semiclassical bulk spacetime. By subregion duality \cite{subregion1, subregion2}, the reduced density matrices (insofar as they are well defined, see \cite{blancocasini, anec, araki76, haagbook} for a discussion of subtleties)
\begin{equation}
\rho_A = {\rm Tr}_{\bar{A}}\, \rho \qquad {\rm and} \qquad \rho_B = {\rm Tr}_{\bar{B}}\, \rho
\end{equation}
describe the physics in entanglement wedges $E(A)$ and $E(B)$, which are bounded by Ryu-Takayanagi (RT) surfaces $a(A)$ and $a(B)$. The two reduced states also define two modular Hamiltonians $H^{\rm mod}_A$ and $H^{\rm mod}_B$ according to $\rho_A = e^{-H^{\rm mod}_A}$. In case of subtleties, we may choose to work with two-sided modular Hamiltonians 
\begin{equation}
H^{\rm two-sided}_A = H^{\rm mod}_A \otimes 1_{\bar{A}} - 1_A \otimes H^{\rm mod}_{\bar{A}}\,,
\label{def2sided}
\end{equation} 
which are always well-defined \cite{araki76, haagbook}.

Modular parallel transport \cite{sewingkit} is generated by a solution $V d\lambda$ of the equation:
\begin{align}
[V d\lambda, H^{\rm mod}_A] & = H^{\rm mod}_B - H^{\rm mod}_A + \textrm{(terms that commute with $H^{\rm mod}_A$)} \nonumber \\
& = dH^{\rm mod}_R \big|_\lambda + (\ldots)
\label{mpt1}
\end{align}
We write $V d\lambda$ as an operator-valued differential form because the right-hand-side is an operator-valued differential form. 

As stated, solutions of (\ref{mpt1}) are ambiguous by the addition of terms, which commute with $H^{\rm mod}_A$. (From here on, we will refer to such terms as zero modes of $H^{\rm mod}_A$.) Modular parallel transport is singled out by a second condition, which stipulates that $V$ contain no zero modes:
\begin{equation}
P^0_A [V] = 0
\label{mpt2}
\end{equation}
In this equation, $P^0_A [\ldots]$ is a projector onto zero modes, which we assume can be consistently defined. \bartek{As is explained and illustrated in Reference~\cite{virasoroberry}, this assumption is subtle because the relevant operators $V$ are not guaranteed to be bounded. In such a context, necessary and sufficient conditions under which projector $P^0_A[ \ldots]$ exists---as well as its uniqueness and explicit form---present challenging problems, which fall outside the scope of this paper. To make progress, we will assume that $P^0_A [\ldots]$ exists and ignore the question of its uniqueness.} 
%

As discussed in Reference~\cite{sewingkit} (see also \cite{modberry}), conditions~(\ref{mpt1}-\ref{mpt2}) define a natural analogue of the Berry connection $V d\lambda \sim -i \langle \psi | d\psi \rangle$, which computes an additional phase that arises when a system in the ground state $|\psi\rangle$ is subjected to an adiabatically changing Hamiltonian \cite{berryref}. Here we change modular Hamiltonians instead of the dynamical Hamiltonian and vary their `thermal states' $e^{-H^{\rm mod}_A} = \rho_A$ rather than the ground states, but the underlying principles are the same. For this reason, holonomies produced by integrating $V d\lambda$ have been called `modular Berry phases.'

For the purposes of the present paper, it is useful to inspect the bulk realization of $V d\lambda$. The JLMS relation identifies the modular Hamiltonians $H^{\rm mod}_R$ in (\ref{mpt1}) with vector fields, which generate boosts orthogonal to the RT surfaces $a(R)$ \cite{jlmsref}. This identification is valid to leading order in the holographic parameter $1/N$ in a neighborhood of the RT surface, but gets modified by non-local terms further away from it. Equation~(\ref{mpt1}) finds a boundary operator $V$, whose bulk action maps one RT surface to another, as well as mapping their orthogonal boosts. Reference~\cite{sewingkit} identified $V$ with a bulk vector field with the same properties.

Condition~(\ref{mpt2}) specifies a unique such vector field. The ambiguity fixed by (\ref{mpt2}) concerns terms, which commute with $H^{\rm mod}_A$. In the neighborhood of the RT surface $a(A)$ and at leading order in $1/N$, they are $a(A)$-preserving diffeomorphisms \cite{sewingkit}. Equation~(\ref{mpt2}) therefore demands that the bulk vector field, which operator $V$ realizes in the boundary theory, map the RT surface $a(A)$ to $a(B)$ without involving $a(A)$-preserving diffeomorphisms. The latter comprise longitudinal diffeomorphisms that act within the RT surface, as well as orthogonal boosts generated by $H^{\rm mod}_A$ itself. 

\paragraph{Special case: Rotation without slipping}
This example was discussed in detail in \cite{modberry}. Consider a differentiable spacelike curve $\mathcal{C}$ in a 2+1-dimensional bulk spacetime. The differentiability assumption means that the curve has a well-defined tangent at every point. Under suitable assumptions,\footnote{In spacetimes other than pure AdS$_3$, the curve must not lie too deep in the bulk and must not be `too radial' anywhere; see \cite{earlylampros, entwinement} for a more complete discussion. \label{ftradial}} the geodesics tangent to $\mathcal{C}$ can be assumed to satisfy the conditions stipulated by the HRT proposal. If so, they define a family of boundary regions whose RT surfaces $a(R(\lambda))$ are tangent to $\mathcal{C}$. 

\begin{figure}[t]
\centering
\includegraphics[width=.80\textwidth]{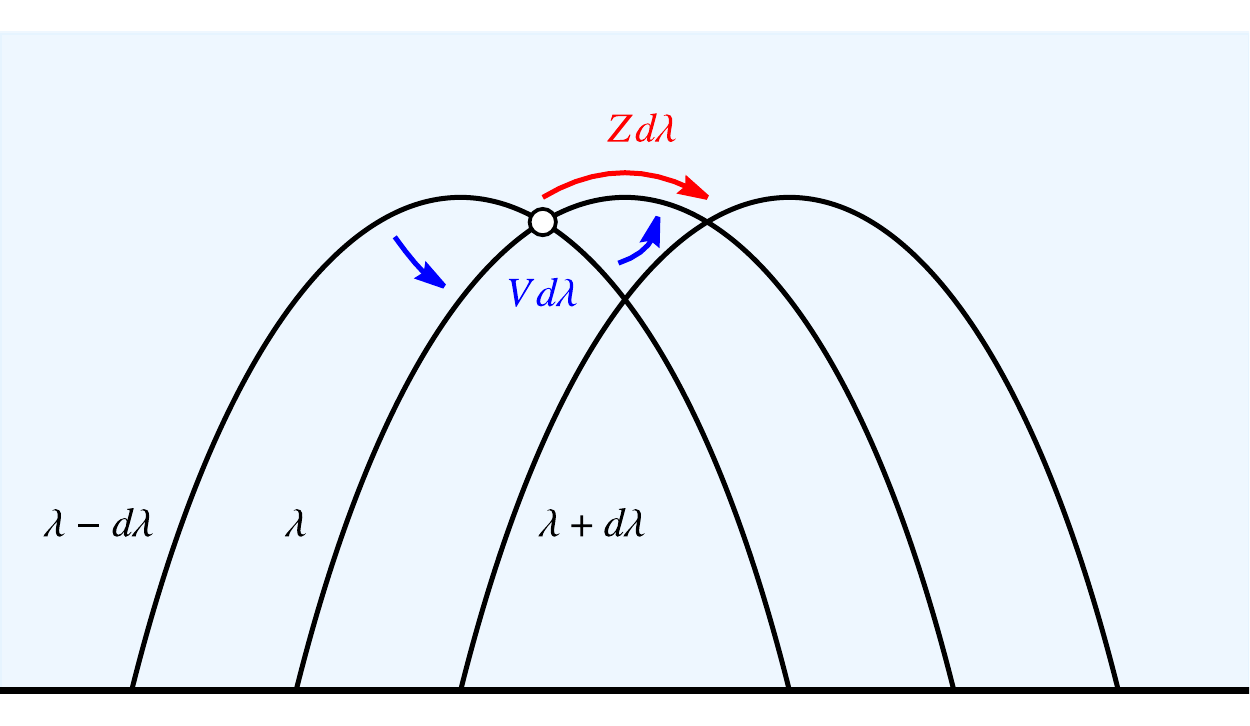}
\caption{Modular parallel transport, which is generated by $V d\lambda$, is a bulk rotation in the neighborhood of a point where two RT surfaces intersect. In a setup with time reflection symmetry, the operator $Z d\lambda$ is a translation between two such intersection points.} 
\label{fig:MPT}
\end{figure}

We now specialize to the case where the curve $\mathcal{C}$ lives on a time reflection-symmetric slice of the bulk spacetime. Under this assumption, the geodesics $a(R(\lambda))$ and $a(R(\lambda + d\lambda))$ intersect in the bulk and the curve can be approximated as a polygon, up to $\mathcal{O}(d\lambda^2)$ corrections. The polygon consists of line segments, which connect intersection points $a(R(\lambda - d\lambda)) \cap a(R(\lambda))$ and $a(R(\lambda)) \cap a(R(\lambda + d\lambda))$ of consecutive geodesics. Modular parallel transport along the kinematic space trajectory induced by $\mathcal{C}$ maps $a(R(\lambda))$ to $a(R(\lambda + d\lambda))$ without using $a(R(\lambda))$-preserving diffeomorphisms. Near their intersection point, this geometric transformation is a rotation by the angle between the two tangent geodesics $a(R(\lambda))$ and $a(R(\lambda + d\lambda))$. It equals $K d\lambda$, where $K$ is the extrinsic curvature of $\mathcal{C}$. In summary, we find that modular parallel transport reduces to a rotation by $K d\lambda$ about a point on the curve. Geometrically, we have described how a straight line rotates while maintaining tangency to curve $\mathcal{C}$, hence `rotation without slipping.' The conclusion is illustrated in Figure~\ref{fig:MPT}.

In the geometric problem posed in equation~(\ref{ourproblem}), the generator of modular parallel transport plays the role of $V d\lambda$. It is the extra rotation, which---when combined with translation $Z d\lambda$ along $\mathcal{C}$---generates the flat transport $A d\lambda$ exemplified by the GPS analogy. When discussing that example, we stated that the extra rotation employed by the navigation unit is fixed by the extrinsic curvature of the trajectory. We have just reached the same conclusion for the generator of modular parallel transport. 

\paragraph{Modular scrambling modes}
Equations~(\ref{mpt1}-\ref{mpt2}), which define modular parallel transport, involve the adjoint action of the modular Hamiltonian. It is useful to think of $[\ldots, H^{\rm mod}_A]$ as a superoperator (sometimes called the Liouvillian) and{\bartek{---if possible\footnote{\bartek{Once again, complications discussed in \cite{virasoroberry} mean that this assumption is subtle and does not hold under most general circumstances. Our final conclusions only require the existence of projectors $P^0[\ldots]$ (equation~\ref{mpt2}) and  $P^\pm[\ldots]$ (equation~\ref{scramblingprojs}), which is a far milder assumption.}}---}}organize operators of the theory into its eigenspaces: 
\begin{equation}
[\mathcal{O}_\nu, H_A^{\rm mod}] = \nu \mathcal{O}_\nu
\label{modmodes}
\end{equation}
If $dH^{\rm mod}$ on the right hand side of (\ref{mpt1}) can be expanded in the basis (\ref{modmodes})
\begin{equation}
dH^{\rm mod}_R\big|_\lambda = d\lambda \sum_\nu c^\nu \mathcal{O}_\nu
\end{equation}
then we can immediately write down the form of modular parallel transport:
\begin{equation}
V = \sum_{\nu \neq 0} \frac{c^\nu}{\nu} \mathcal{O}_\nu
\label{mptformal}
\end{equation}
The eigenvalues $\nu$ in (\ref{modmodes}) are called modular frequencies. Reference~\cite{modularchaos} showed that they obey the \emph{modular chaos bound}:
\begin{equation}
-2\pi \leq {\rm Im}\,\nu \leq 2\pi 
\label{chaosbound}
\end{equation}
The bound encapsulates chaotic properties of modular flow, which are necessary for the emergence of Poincar{\'e} symmetry in the bulk. 

The imaginary component of a modular frequency plays a crucial role in bulk reconstruction. This quantity has no analogue in finite-dimensional quantum mechanics. To see this, write the modular Hamiltonian as a diagonal matrix with eigenvalues $E_i$ and recognize that all eigenmodes of $[\ldots, H^{\rm mod}]$ are matrices with a single non-zero entry, whose modular frequencies are $\nu = E_i - E_j$. This argument does not apply in infinite-dimensional systems, where the imaginary component of $\nu$ gains a crucial significance. 

As explained in \cite{modularchaos}, bound~(\ref{chaosbound}) is necessarily saturated in holographic theories of quantum gravity. In this text we follow \cite{modularchaos} in assuming that the modular frequencies that saturate (\ref{chaosbound}) are purely imaginary. Violating this assumption would complicate several of our formulas but not invalidate our reasoning; we decided to adopt the assumption for the sake of clarity because no counterexamples are known. Solutions of the eigenvalue equation
\begin{equation}
[G_\mu, H^{\rm mod}_{R(\lambda)}] = 2\pi i G_\mu
\qquad {\rm and} \qquad
[G_{\bar{\mu}}, H^{\rm mod}_{R(\lambda)}] = - 2\pi i G_{\bar\mu}
\label{scramblingmodes}
\end{equation}
are modular scrambling modes. We leave the index $\mu$ unbarred for scrambling modes with modular frequency $+2\pi i$ and put an overbar on those with frequency $-2 \pi i$. The index $\mu$ (respectively $\bar{\mu}$) will distinguish scrambling modes of the same frequency if there are more than one. Whenever the distinction between $\mu$ and $\bar{\mu}$ is immaterial, we will use the Latin index $m$, e.g. $G_m$. 

In the bulk, in the neighborhood of the RT surface $a(A)$, the adjoint action of the $G_m$ is that of vector fields, which generate null deformations of $a(A)$. For this reason, modular scrambling modes control the geometric action of the generator of modular parallel transport $V$ defined in (\ref{mpt1}). Although in general the form $V d\lambda$ may involve both modular scrambling modes and other modes $\mathcal{O}_\nu$, it is the former that encapsulate the geometric action of modular parallel transport in the bulk spacetime.

\paragraph{Example: pure AdS$_3$} An explicit example of modular scrambling modes is supplied by the vacuum of a 1+1-dimensional conformal field theory on a plane. We denote CFT space as $x$ and time as $t$. Without loss of generality, we consider interval $x \in (L, R)$ on an equal time slice. Then conformal symmetry fixes the form of the modular Hamiltonian:
\begin{align}
H^{\rm mod} & = H^{\rm mod}_{\rm left} + H^{\rm mod}_{\rm right} \nonumber \\
& \equiv \frac{2 \pi i}{R - L} \big( L_+ - (L + R) L_0 + (L R)\, L_- \big) + \big(L_{...} \leftrightarrow \bar{L}_{...}\big)
\label{hmod}
\end{align}
Here $L_\pm, L_0$ are generators of the left-moving component of the global conformal group $SL(2,\mathbb{R})$ and $\bar{L}_\pm, \bar{L}_0$ are their right-moving counterparts. The modular Hamiltonian of the vacuum splits into a left-moving and right-moving component. It is easy to see that (\ref{hmod}) fixes points $x = L, R$ and its Euclidean continuation acts as a clockwise rotation near $x = R$ (anticlockwise near $x = L$).

As a consequence of the half-sided modular inclusion theorem \cite{modularinclusion, modinclcasini, samleut}, the modular scrambling modes take on a particularly simple form \cite{modularchaos}:
\begin{align}
\tilde{G}_{\rm left} & = \frac{\partial H^{\rm mod}_{\rm left}(R, L)}{\partial L}
= + \frac{2\pi i}{(R-L)^2} \big( L_+ - 2 R\, L_0 + R^2 L_-\big)  
\\
\tilde{G}_{\overline{\rm left}} & = \frac{\partial H^{\rm mod}_{\rm left}(R, L)}{\partial R} 
= - \frac{2\pi i}{(R-L)^2} \big( L_+ - 2 L\, L_0 + L^2 L_- \big)
\end{align}
In the right-moving sector, these relations are reversed:
\begin{equation}
\tilde{G}_{\rm right} = \frac{\partial H^{\rm mod}_{\rm right}(R, L)}{\partial R}
\qquad {\rm and} \qquad
\tilde{G}_{\overline{\rm right}} = \frac{\partial H^{\rm mod}_{\rm right}(R, L)}{\partial L}
\end{equation}
Here the index $\mu$ from equation~(\ref{scramblingmodes}) ranges over $\mu =$~left, right; the same applies to $\bar{\mu}$. We can associate $\mu$ with left- and right-movers in this way because of the split~(\ref{hmod}) in the modular Hamiltonian of the vacuum. In general states, however, $\mu$ will be an arbitrary label. Finally, to remind the reader of our notation, we reiterate that the collective index $m$ ranges here over $m = $~left, right, $\overline{\rm left}$, $\overline{\rm right}$.

A convenient normalization for these modular scrambling modes is the following:
\begin{align}
G_{\overline{\rm left}} & = \phantom{\big(}L_+ - 2 L\, L_0 + L^2 L_- \\
G_{\rm left} & = \big(L_+ - 2 R\, L_0 + R^2 L_- \big) / (R-L)^2
\end{align}
Then the three operators satisfy the canonical commutation relations of the $SL(2,\mathbb{R})$ algebra under the identification:
\begin{equation}
- \frac{H^{\rm mod}_{\rm left}}{2\pi i} \leftrightarrow L_0 \qquad 
G_{\rm left} \leftrightarrow L_{-} \qquad 
 G_{\overline{\rm left}} \leftrightarrow L_{+}
\label{sl2ridentification}
\end{equation}
This identification is an exact equality on the half-line $(L,R) = (0, \infty)$. 

For motions in the vacuum kinematic space, the form of the modular scrambling modes and equation~(\ref{mptformal}) immediately give the generator of modular parallel transport:
\begin{equation}
V d\lambda = \frac{d\lambda}{2\pi i} 
\left( \frac{\partial R}{\partial \lambda} \Big( \tilde{G}_{\rm right} - \tilde{G}_{\overline{\rm left}} \Big) 
+ \frac{\partial L}{\partial \lambda} \Big( \tilde{G}_{\rm left} - \tilde{G}_{\overline{\rm right}} \Big)\right)
\label{vdifferentials}
\end{equation}
In holographic theories, the four $\tilde{G}_m$ act in the bulk as global isometries of AdS$_3$. The remaining two global isometries of AdS$_3$ are zero modes of $H^{\rm mod}$. 

\section{A new transport problem}
\label{sec:maingen}
The previous section introduced the generator of modular parallel transport $V d\lambda$. Every instance of the modular transport problem is specified by a trajectory in kinematic space, which we parameterize by $\lambda$. 

This section posits a new transport problem, which is related to but distinct from modular parallel transport. It, too, takes as input a trajectory in kinematic space parameterized by $\lambda$. We will denote the generator of the new type of transport $Z d\lambda$. At first, the transport problem solved by $Z d\lambda$ may seem somewhat academic. However, when we specialize to three bulk dimensions in the next section, we will find that $Vd\lambda + Zd\lambda \equiv A d\lambda$ generates the type of GPS-like transport described in the Introduction. One further step, which is explained in Section~\ref{sec:notimerefl}, relates $Z d\lambda$ to the boundary generator of bulk parallel transport along a spacelike curve.

We start by relating modular parallel transport to the apparatus of differential geometry.

\subsection{Kinematic tangent space}
Consider a trajectory in kinematic space, here understood as the space of boundary regions. Modular parallel transport in kinematic space is generated by $V d\lambda$, which solves equations~(\ref{mpt1}-\ref{mpt2}). In general, the operator-valued form $V d\lambda$ combines scrambling modes~(\ref{scramblingmodes}) and other modes of $[\ldots, H^{\rm mod}]$, but it is the former that describe motions of RT surfaces in the bulk spacetime.\footnote{\bartek{We assume that scrambling modes are not involved in state deformations, which leave the boundary region $R(\lambda)$ unchanged. This is obvious for state deformations that preserve the entanglement spectrum of $R(\lambda)$ because the action of scrambling modes is non-unitary. For spectrum-changing deformations, the assumption is plausible but non-trivial. We are not aware of counterexamples.} \label{ftstatescr}} Because our motivation is to study bulk vector fields, we drop non-scrambling terms in $V d\lambda$ and focus attention on:
\begin{equation}
\tilde{V} d\lambda \equiv P^+ [V]\, d\lambda + P^- [V]\, d\lambda
\label{deftildev}
\end{equation}
Here $P^{\pm}[\ldots]$ are projectors onto modular scrambling modes of frequency $\pm 2\pi i$, which are formally given by \cite{modularchaos}:
\begin{equation}
P^{\pm} [V] = \lim_{\Lambda \to \pm \infty} \frac{1}{2\Lambda} \int_{-\Lambda}^\Lambda ds\,
e^{\mp 2 \pi s} e^{-i s H^{\rm mod}} V e^{i s H^{\rm mod}}    
\label{scramblingprojs}
\end{equation}
Let us fix a local basis of modular scrambling modes around any point (boundary region) $\lambda$ in kinematic space. This means a choice of indexing for scrambling modes with labels $\mu$ and $\bar{\mu}$, as discussed around equation~(\ref{scramblingmodes}). In the notation defined there, the scrambling part of the modular parallel transport generator can be written as:
\begin{equation}
\tilde{V} d\lambda = 
G_\mu \frac{dx^\mu}{d\lambda} d\lambda + G_{\bar{\mu}} \frac{dx^{\bar{\mu}}}{d\lambda} d\lambda
\equiv G_m \frac{dx^m}{d\lambda} d\lambda
\label{tvgxl}
\end{equation}
in terms of local coordinates $x^\mu$ and $x^{\bar{\mu}}$ (collectively $x^m$) in kinematic space. (We are using Einstein's summation convention on matching upper and lower indices.) By definition, the modular scrambling modes generate local kinematic space translations in these coordinates:
\begin{equation}
G_\mu \leftrightarrow \partial_\mu
\qquad {\rm and} \qquad
G_{\bar{\mu}} \leftrightarrow \partial_{\bar{\mu}}
\label{tks}
\end{equation}
Effectively, we are identifying the span of scrambling modes~(\ref{scramblingmodes}) with the tangent space $T_\lambda$ at point $\lambda$ in kinematic space.

\paragraph{Kinematic Christoffel symbols}
We would like to understand how these generators transform under infinitesimal translations. Consider a motion between two neighboring points $\lambda$ and $\lambda + d\lambda$ in kinematic space. 
It is sufficient to consider the case where $d\lambda = dx^m$. We take scrambling modes $G_\mu$ (respectively $G_{\bar{\mu}}$), which satisfy 
\begin{equation}
[G_\mu(\lambda), H^{\rm mod}(\lambda)] = 2\pi i G_\mu (\lambda)
\qquad {\rm and} \qquad
[G_{\bar{\mu}}(\lambda), H^{\rm mod}(\lambda)] = -2\pi i G_{\bar{\mu}}(\lambda)
\end{equation}
and compare them with the scrambling modes at $\lambda + d\lambda = \lambda + dx^m$. The change in $H^{\rm mod}$ induced by $G^m$ is given by equation (\ref{mpt1}), so we have
\begin{align}
H^{\rm mod}(\lambda + dx^m) 
& = H^{\rm mod}(\lambda) + dx^m [G_m(\lambda), H^{\rm mod}(\lambda)] \\
& = H^{\rm mod}(\lambda) \pm 2\pi i G_m(\lambda) dx^m
\qquad \textrm{(no summation)} \nonumber
\end{align}
The sign depends on the frequency of the scrambling mode, that is whether $G_m = G_\mu$ or $G_m = G_{\bar{\mu}}$. We then take the scrambling modes of $H^{\rm mod}(\lambda + dx^m)$, also labeled by $\mu$ and $\bar{\mu}$. In other words, we solve
\begin{align}
[G_\mu(\lambda+dx^m), H^{\rm mod}(\lambda+dx^m)] & = \phantom{-} 2\pi i G_\mu (\lambda+dx^m) \\
[G_{\bar{\mu}}(\lambda+dx^m), H^{\rm mod}(\lambda+dx^m)] & = -2\pi i G_{\bar{\mu}}(\lambda+dx^m)
\nonumber
\end{align}
We assume that the labels $\mu$ and $\bar{\mu}$ at kinematic points $\lambda$ and $\lambda + d\lambda$ are chosen such that the operators $G_\mu$ and $G_{\bar{\mu}}$ are continuous functions of $\lambda$. We can then take a partial derivative of $G_\mu$ (respectively $G_{\bar{\mu}}$) in the $m$-direction, and expand it in modes of $H^{\rm mod}(\lambda)$. Focusing on scrambling modes, we define:
\begin{equation}
\frac{\partial G_\nu}{\partial x^m} 
= \Gamma^{\sigma}_{m\nu} G_\sigma + \textrm{non-scrambling} 
\qquad {\rm and} \qquad
\frac{\partial G_{\bar{\nu}}}{\partial x^m}
= \Gamma^{\bar{\sigma}}_{m\bar{\nu}} G_{\bar{\sigma}} + \textrm{non-scrambling~~}
\label{kschristoffels}
\end{equation}
The coefficients $\Gamma^{\sigma}_{m\nu}$ and $\Gamma^{\bar{\sigma}}_{m\bar{\nu}}$ are kinematic Christoffel symbols. \bartek{We emphasize that this eponym does not implicate the existence of a metric on kinematic space. The Christoffel symbols~(\ref{kschristoffels}) define a connection, not necessarily a metric-compatible connection.}

Definition~(\ref{kschristoffels}) follows how Christoffel symbols are typically defined in the context where the tangent spaces over all points can be embedded in one common vector space. A canonical example is the construction of Christoffel symbols over $S^d$, which is embedded in $\mathbb{R}^{d+1}$. There, one can extend the action of a vector field in $T_\lambda S^d$ to $T_\lambda \mathbb{R}^{d+1}$, then identify $T_\lambda \mathbb{R}^{d+1}$ with $T_{\lambda + d\lambda} \mathbb{R}^{d+1}$ and project down from $T_{\lambda + d\lambda} \mathbb{R}^{d+1}$ to $T_{\lambda + d\lambda} S^d$. In a similar fashion, we embed the scrambling modes at $\lambda$ in the space of all operators in the theory, then project down to the scrambling modes at $\lambda + d\lambda$. The coefficients in the resulting expansion $P^+[\partial G_\nu / \partial x^m] = \Gamma^{\sigma}_{m\nu} G_\sigma$ (and likewise for the barred modes) are kinematic Christoffel symbols.

Before continuing, we remark that all Christoffel symbols that mediate between $+2\pi i$ and $-2\pi i$ eigenmodes vanish: $\Gamma^\sigma_{m \bar{\nu}} = 0 = \Gamma^{\bar{\sigma}}_{m \nu}$. This fact kills off half of the potential components of $\Gamma^{p}_{mn}$. To confirm it, apply $\partial_m$ to the defining equation of the scrambling mode:
\begin{equation}
\partial_m [G_{\bar{\nu}}, H^{\rm mod}] 
= [\partial_m G_{\bar{\nu}}, H^{\rm mod}] + [G_{\bar{\nu}}, \partial_m H^{\rm mod}]
= - 2\pi i \partial_m G_{\bar{\nu}} 
\label{partialtodef}
\end{equation}
A nonvanishing $\Gamma^\sigma_{m \bar{\nu}}$ would be read off from (\ref{partialtodef}) by projecting onto the $(+2\pi i)$-eigenspace of $[\ldots, H^{\rm mod}]$:
\begin{equation}
\!\!\! [P^+[\partial_m G_{\bar{\nu}}], H^{\rm mod}] + P^+[[G_{\bar{\nu}}, \partial_m H^{\rm mod}]] 
\stackrel{*}{=} + 2\pi i P^+[\partial_m G_{\bar{\nu}}] + 0
= - 2\pi i P^+[\partial_m G_{\bar{\nu}}]\,\,\,
\label{whysomegammais0}
\end{equation}
In the step marked with a star, we apply the definition of $P^+[\ldots]$ to the first term and observe that the second term vanishes. This is because a $(+2\pi i)$-eigenvalue component of $[G_{\bar{\nu}}, \partial_m H^{\rm mod}]$ can only come from a $(+4\pi i)$-eigenvalue component of $\partial_m H^{\rm mod}$, but those are forbidden by the modular chaos bound \cite{modularchaos}. 

\subsection{The transport problem}
\paragraph{Rough idea}
We continue to consider a trajectory in kinematic space parameterized by $\lambda$. We can use kinematic coordinates defined in~(\ref{tvgxl}) to locally express $\tilde{V} d\lambda$ in terms of partial `velocities' $dx^m / d\lambda$. Using the expansion~(\ref{tvgxl}) and the kinematic Christoffel symbols~(\ref{kschristoffels}), we define the covariant derivative of $\tilde{V}$:
\begin{equation}
D_\lambda \tilde{V} \equiv 
\frac{d^2x^m}{d\lambda^2} G_m + \Gamma^p_{mn}\frac{dx^m}{d\lambda} \frac{dx^n}{d\lambda} G_p
\label{dvdef}
\end{equation}
Note that this definition excludes terms, which come from the non-scrambling modes in (\ref{kschristoffels}). This allows us to think of $D_\lambda \tilde{V}$ as a vector in the tangent space $T_\lambda$ to kinematic space, which is spanned by scrambling modes. We can interpret~(\ref{dvdef}) as the covariant acceleration of a trajectory in kinematic space. 

With~(\ref{tvgxl}) and (\ref{dvdef}) in place, we are ready to sketch the new transport problem. In spirit, we are looking for an operator-valued form $Z d\lambda$, which satisfies:
\begin{equation}
[\tilde{V}, [\tilde{V}, Zd\lambda]] = [\tilde{V}, D_\lambda \tilde{V} d\lambda]
\qquad\qquad \textrm{(rough idea)}
\label{defzcartoon}
\end{equation}
We explain the rigorous condition that selects $Zd\lambda$ momentarily. Before introducing complications, however, let us comment on the qualitative meaning of (\ref{defzcartoon}). 

\bartek{The right hand side is a commutator of operators, which represent kinematic velocity and acceleration. On the left, we have a commutator of the velocity and another operator} $[\tilde{V}, Z]$ to be solved for. Both $D_\lambda \tilde{V} d\lambda$ and $[\tilde{V}, Z] d\lambda$ represent transformations of $\tilde{V}$. The first term, $D_\lambda \tilde{V} d\lambda$, is the infinitesimal change under translations along the curve in kinematic space---that is, under modular parallel transport. After including an extra minus sign, $-[Z, \tilde{V}]d\lambda$ is the transformation under the adjoint action of $Z$. Condition~(\ref{defzcartoon}) sets the two equal, up to terms which commute with $\tilde{V}$. The idea of (\ref{defzcartoon}) is that under the combined transformation $D_\lambda \tilde{V} + [Z, \tilde{V}]$ the direction of the velocity vector should remain constant. The condition is phrased in terms of commutators $[\tilde{V}, \ldots]$ rather than directly pitting $D_\lambda \tilde{V}$ against $[Z, \tilde{V}]$ because we wish to allow the velocity to change in magnitude; only its direction should be preserved. This is the essence of GPS-style transport, which we sketched in the Introduction: the display always points ahead, even when the car speeds up or slows down. 

\paragraph{Technical details}
There are several reasons why (\ref{defzcartoon}) can only be a cartoon of the full-fledged condition. As a preliminary, we remind the reader that we are identifying the tangent space to kinematic space with the span of scrambling modes (\ref{scramblingmodes}). 
\begin{itemize}
\item In order to think of $-[Z, \tilde{V}]d\lambda$ as an infinitesimal transformation of the vector $\tilde{V}$, we must interpret $[Z, \ldots]$ as a map from tangent space to itself. If $Z$ is a modular zero mode, this is automatic. This follows from
\begin{equation}
[H^{\rm mod}, A] = aA \quad {\rm and} \quad [H^{\rm mod}, B] = bB 
\quad \Longrightarrow \quad [H^{\rm mod}, [A,B]] = (a+b) [A,B],
\end{equation}
which is a consequence of the Bianchi identity. In particular, assuming $Z$ is a zero mode, if $\tilde{V}$ is a linear combination of scrambling modes then so is $[Z, \tilde{V}]$. Accordingly, {\bf we demand that $Z$ be a modular zero mode}. 

Looking ahead, this demand is reasonable vis-{\`a}-vis the anticipated interpretation of $Z$ as a generator of translation along a bulk curve $\mathcal{C}$. There, the modular Hamiltonian in question locally generates a boost, which acts orthogonally to a geodesic tangent to $\mathcal{C}$. Precisely because the boost is orthogonal, the translation along the tangent geodesic commutes with it. 
\item Now each of the three operators $\tilde{V}$, $[\tilde{V}, Z]$, $D_\lambda \tilde{V}$ is a linear combination of scrambling modes. Their commutators will therefore involve modes of frequencies $\pm 4\pi i$ and $0$. The former do not have a geometric interpretation. Accordingly, we limit our attention only to the zero mode component of condition~(\ref{defzcartoon}). In a rigorous treatment, {\bf we apply $P^0$ on both sides of the condition that sets $Z$.}
\item After implementing the previous two points, we are to look for a zero mode $Z$, which the map $P^0 \Big[ [\tilde{V}, [\tilde{V}, \ldots]] \Big]$ sends to $P^0 \Big[ [\tilde{V}, D_\lambda \tilde{V}] \Big]$. Under most general circumstances, the map $P^0 \Big[ [\tilde{V}, [\tilde{V}, \ldots]] \Big]$ might not be invertible. This happens, for example, in the presence of superselection sectors.\footnote{Consider a Hilbert space $\mathcal{H} = \mathcal{H}_1 \oplus \mathcal{H}_2$, and density operator $\rho = p \rho_1 + (1-p) \rho_2$, with $\rho_{1}$ acting on $\mathcal{H}_{1}$ (and $\rho_2$ on $\mathcal{H}_2$). Then kinematic space motion that changes $\rho_1$ commutes with the zero mode $-\log \rho_2$.} We have encountered the same caveat before in the definition of modular parallel transport. Following the same logic, we stipulate that: 
\begin{itemize}
\item[(i)] $Z$ contains no terms, which are annihilated by $P^0 \Big[ [\tilde{V}, [\tilde{V}, \ldots]] \Big]$. 
\item[(ii)] $P^0 \Big[ [\tilde{V}, [\tilde{V}, \ldots]] \Big]$ must reproduce $P^0 \Big[ [\tilde{V}, D_\lambda \tilde{V}] \Big]$ only up to terms, which are annihilated by $P^0 \Big[ [\tilde{V}, [\tilde{V}, \ldots]] \Big]$.
\end{itemize}
In this work we simply assume that the projectors necessary to enforce conditions (i)-(ii) exist, but it would be interesting to study their existence and explicit form. 
\end{itemize}
\smallskip

\paragraph{Summary of transport problem:}
\begin{align}
P^0_{R(\lambda)} \Big[ [\tilde{V}(\lambda), [\tilde{V}(\lambda), Z(\lambda)]] \Big] 
& = P^0_{R(\lambda)} \Big[ [\tilde{V}(\lambda), D_\lambda \tilde{V}] \Big] 
\label{problem1} \\
& ~~~~~ + \textrm{terms annihilated by 
$P^0_{R(\lambda)} \Big[ [\tilde{V}(\lambda), [\tilde{V}(\lambda), \,\ldots]] \Big]$} \nonumber \\
[H^{\rm mod}(\lambda), Z(\lambda)] & = 0 
\label{problem2} \\
Z(\lambda)~\textrm{\bartek{is annihilated by}}&~~~~\left(\textrm{\bartek{Projector onto}}~{\rm ker}\,P^0_{R(\lambda)} \Big[ [\tilde{V}(\lambda), [\tilde{V}(\lambda), \,\ldots]] \Big]\right)
\label{problem3}
\end{align}
\smallskip

\subsection{The kinematic vielbein postulate} 
\label{sec:vielbein}
We would like to situate our transport problem in a broader context. In order not to clutter the discussion, we ignore the subtlety associated with the kernel of $P^0 \Big[ [\tilde{V}, [\tilde{V}, \ldots]] \Big]$ and focus on the zero mode-projected equation~(\ref{defzcartoon}). In components, that equation reads:
\begin{equation}
\!\!\!
P^0 \Big[ [G_m \frac{dx^m}{d\lambda},\, [G_m \frac{dx^m}{d\lambda}, Zd\lambda]] \Big]
= P^0 \Big[ [ G_m \frac{dx^m}{d\lambda},\,
\left( \frac{d^2x^p}{d\lambda^2} + \Gamma^p_{mn}\frac{dx^m}{d\lambda} \frac{dx^n}{d\lambda} \right) G_p d\lambda] \Big]\,\,
\label{finalcond}
\end{equation}
As the $G_m$ define coordinates on the tangent space $T_\lambda$ of kinematic space, the operation $[\ldots, Zd\lambda]$ is a change of basis on the tangent space. This observation gives a useful perspective on equation~(\ref{finalcond}).

In differential geometry, transport of a basis of tangent space is controlled by the spin connection $\omega$. In order to relate equation~(\ref{finalcond}) to a kinematic spin connection, we consider a family of kinematic trajectories, which are indexed by $a$ and parameterized by $\lambda^a$. We then interpret the derivatives $dx^m / d\lambda^a$ as components of a kinematic \bartek{frame field} $e^m_a$. This defines a new tangent space basis $G_a$ via:
\begin{equation}
G_m (dx^m/d\lambda^a) = G_m e^m_a \equiv G_a
\label{defga}
\end{equation}
The spin connection $(\omega_m)^a_b$ then appears in the expansion of the covariant derivative of a kinematic vector field $V = v^a G_a$:
\begin{equation}
\nabla V 
= \big(\partial_m v^a + (\omega_m)^a_b v^b\big) dx^m G_a 
= \big(\partial_m v^a + (\omega_m)^a_b v^b\big) dx^m e^p_a G_p 
\label{defomega}
\end{equation}
We now write the same expression using $V = v^p G_p$:
\begin{align}
\nabla V 
& = \big( \partial_m v^p + \Gamma_{mn}^p v^n \big) dx^m G_p = 
\big( \partial_m (e^p_a v^a) + \Gamma_{mn}^p v^n \big) dx^m G_p
\nonumber \\
& = \big( e^p_a \partial_m v^a + (\partial_m e^p_b) v^b + \Gamma_{mn}^p e^n_b v^b \big) dx^m G_p
\label{2ndexpr} 
\end{align}
Equating the two expressions gives the so-called vielbein postulate (see e.g.~\cite{vielbeinpost}):
\begin{equation}
\nabla_m e^p_b \equiv \partial_m e^p_b + \Gamma_{mn}^p e^n_b - (\omega_m)^a_b e^p_a = 0
\label{thepostulate}
\end{equation}
\bartek{Strictly speaking, our $e_a^m$ is not yet a vielbein because we have no kinematic metric to set $e_a^m e_b^n g_{mn} = \eta_{ab}$. Nevertheless, we will refer to equation~(\ref{thepostulate}) as the `vielbein postulate' to follow common parlance.}

\paragraph{Relation to the transport problem} Equation~(\ref{finalcond}) pertains to one trajectory in kinematic space, which we can take to be parameterized by one $\lambda^a$. The vielbein postulate (\ref{thepostulate}) guarantees that defining an operator-valued form $Z_a d\lambda^a$ via
\begin{equation}
[G_m, Z_a] \equiv (\omega_m)_a^b\, G_b = (\omega_m)_a^b e_b^p\, G_p 
\equiv (\omega_m)_a^p \, G_p
\label{zomega}
\end{equation}
automatically solves condition~(\ref{finalcond}). Indeed, in the notation introduced above, equation~(\ref{finalcond}) takes the form:
\begin{equation}
\big( \partial_m e^p_a + \Gamma^p_{mn} e^n_a  - (\omega_m)_a^p \big) \,
P^0 \Big[ [  e^m_a G_p,\, G_a ]\Big] = 0
\label{condpostulate}
\end{equation}
There is no summation over the index $a$, which appears in three distinct places. We conclude that equation~(\ref{finalcond}) is a weakening of the vielbein postulate. It differs from the vielbein postulate in two important ways. 

First, the vielbein postulate relies on a full set of trajectories indexed by $a$, which together form a \bartek{frame field}. It is sensitive to all those trajectories collectively. (We derived it by selecting the multiple of $v^b dx^m G_p$ in the difference of equations~(\ref{defomega}) and (\ref{2ndexpr}).) In contrast, equation~(\ref{finalcond}) makes sense when only one trajectory---parameterized by a single $\lambda$---is available. Note that solution~(\ref{zomega}) does not involve other indices $a$, as is evident by writing $[G_m, Z_a] = (\omega_m)_a^p \, G_p$. But when a full kinematic \bartek{frame field} is available, we can write solution~(\ref{zomega}) using \bartek{it} and the spin connection: $[G_m, Z_a] = (\omega_m)_a^b e_b^p\, G_p$. 

The second, bigger difference, is manifested by rewriting~(\ref{finalcond}) in the form~(\ref{condpostulate}). Unlike in the vielbein postulate, we do not require the parenthesis to vanish identically, but only when contracted with the zero mode projection of $[e_a^m G_p, G_a]$. This is because $(\omega_m)^b_a$ does not transform as a tensor and, as a result, it does not change homogeneously under rescaling of $\lambda^a$. Only when $\lambda^a$ is appropriately scaled is it possible to write its action in the form~(\ref{zomega}). In this way, equation~(\ref{finalcond}) and its rewriting~(\ref{condpostulate}) are a reparameterization-independent counterpart of the vielbein postulate. We return to this point in Section~\ref{sec:ads3}, where we discuss an example in AdS$_3$, as well as in Applications. 

\section{Bulk parallel transport along spacelike curves}
\label{sec:main}

In the remainder of the paper, we specialize to the AdS$_3$/CFT$_2$ correspondence.

We would like to relate the operator-valued one-form $Z d\lambda$, which solves problem~(\ref{problem1}-\ref{problem3}), to parallel transport along a differentiable spacelike curve $\mathcal{C}$ in the bulk. We work in the setup considered in paragraph `Special case: Rotation without slipping.' The tangents to curve $\mathcal{C}$ define a family of geodesics, which we assume satisfy the conditions of the HRT proposal. The geodesics, in turn, define a family of boundary regions $R(\lambda)$ and modular Hamiltonians $H^{\rm mod}(\lambda)$. The tangent geodesics, which are RT surfaces for $R(\lambda)$, are denoted $a(R(\lambda))$. In this way, curve $\mathcal{C}$ in the bulk induces a curve in kinematic space.

For simplicity, we initially assume that the curve $\mathcal{C}$ lives on a time reflection-symmetric slice of the bulk geometry. We remove that assumption in Section~\ref{sec:notimerefl}.

\subsection{Curves on a time reflection-symmetric bulk slice}
\label{sec:mainrefl}
Under the assumption of time reflection symmetry, modular parallel transport along the kinematic space curve that is induced by a bulk curve $\mathcal{C}$ was analyzed in Section~\ref{sec:rev}. We found that $V(\lambda)$---the generator of modular parallel transport---effects a rotation about the intersection point of RT surfaces $a(R(\lambda))$ and $a(R(\lambda + d\lambda))$. In general, this geometric action can be accompanied by a non-geometric component, which is generated by non-scrambling modes in $V(\lambda)$. To focus on geometric transformations, we follow Section~\ref{sec:maingen} and work with $\tilde{V}(\lambda)$ defined in equation~(\ref{deftildev}). 

We now consider two $\tilde{V}$s, which act at infinitesimally separated values of $\lambda$:
\begin{align}
\tilde{V}(\lambda - d\lambda) & \quad \rightarrow \quad 
\textrm{rotation about}~a(R(\lambda - d\lambda)) \cap a(R(\lambda)) \nonumber \\
\tilde{V}(\lambda) & \quad \rightarrow \quad
\textrm{rotation about}~a(R(\lambda)) \cap a(R(\lambda + d\lambda)) \nonumber
\end{align}
The interpretations on the right are understood to hold in a small neighborhood of the relevant RT surfaces.

Rotations about two different centers do not commute. But we can have 
\begin{align}
[e^{- d\lambda Z} \tilde{V}(\lambda - d\lambda) e^{d\lambda Z}, \tilde{V}(\lambda)] & \approx \nonumber \\
[(1 - d\lambda Z) \tilde{V}(\lambda - d\lambda) (1 + d\lambda Z), \tilde{V}(\lambda)] & = 0
\label{finitecond}
\end{align}
if one rotation gets conjugated by $Z d\lambda$, which translates one center of rotation to the other. Here we write down the translation as an operator-valued one-form because the translation in question is infinitesimal. In particular, if $Z d\lambda$ that satisfies~(\ref{finitecond}) can be found, we will interpret it as a translation from bulk point $a(R(\lambda - d\lambda)) \cap a(R(\lambda))$ to bulk point $a(R(\lambda)) \cap a(R(\lambda + d\lambda))$. This interpretation should be valid in a neighborhood of the RT surface $a(R(\lambda))$. The translation we are after is an $a(R(\lambda))$-preserving diffeomorphism and therefore a zero mode of $H^{\rm mod}(\lambda)$; this agrees with equation~(\ref{problem2}). 

A sequence of such $Z d\lambda$s, taken over a discretized set of $\lambda$s spaced by $d\lambda$, generates a translation along a polygonal approximation to $\mathcal{C}$. This polygonal approximation differs from actual $\mathcal{C}$ at order $d\lambda^2$. In the continuum limit, therefore, $Z d\lambda$ generates bulk parallel transport along $\mathcal{C}$. The construction is illustrated in Figure~\ref{fig:MPT}.

To relate~(\ref{finitecond}) to problem~(\ref{problem1}-\ref{problem3}), we Taylor-expand $\tilde{V}(\lambda - d\lambda)$ to obtain
\begin{equation}
\!\!\!\!\!
[(1 - d\lambda Z) (\tilde{V}(\lambda) - d\lambda D_\lambda \tilde{V}) (1 + d\lambda Z), 
\tilde{V}(\lambda)] \approx
d\lambda\, [[\tilde{V}(\lambda), Z] - D_\lambda \tilde{V}, \tilde{V}(\lambda)]  = 0
\label{dzequal}
\end{equation}
at leading order in $d\lambda$. This is nothing but equation~(\ref{defzcartoon}). As before, we impose equation~(\ref{dzequal}) only at the level of zero modes because other potential contributions to it are not geometric. Furthermore, if there are zero modes which are annihilated by $[[\tilde{V}, [\tilde{V}, \ldots]]$ then we do not include them in the solution $Z d\lambda$. Such terms, if they exist, do not help in relating $V(\lambda)$ to $V(\lambda - d\lambda)$, so they have nothing to do with bulk translations. These two provisos recover equations~(\ref{problem1}) and (\ref{problem3}). 

\subsubsection{Example: Curve in Poincar{\'e}-AdS$_3$}
\label{sec:ads3}
We consider a curve $\mathcal{C}$ on a time reflection-symmetric slice with metric:
\begin{equation}
ds^2 = \frac{dx^2 + dz^2}{z^2}
\end{equation}
The curve is specified as an envelope of bulk geodesics $a(R(\lambda))$, which subtend intervals $(L(\lambda), R(\lambda))$ on the asymptotic boundary. Time reflection symmetry implies that the operator-valued form $Z d\lambda$ must be an (infinitesimal) multiple of the `modular momentum' operator:
\begin{equation}
P^{\rm mod} = \frac{1}{2\pi i} \left( H^{\rm mod}_{\rm left} - H^{\rm mod}_{\rm right} \right)
\label{defpmod}
\end{equation}
Note the relative minus sign, which distinguishes $P^{\rm mod}$ from $H^{\rm mod}$. This is the only operator in the algebra of $SL(2, \mathbb{R}) \times SL(2, \mathbb{R})$, which commutes with $H^{\rm mod}$ and respects time reflection.

Comparing with equation~(\ref{vdifferentials}), the generator of modular parallel transport is:
\begin{equation}
V = \frac{dL}{d\lambda} \partial_L P^{\rm mod} - \frac{dR}{d\lambda} \partial_R P^{\rm mod} = \tilde{V}
\label{vexpl}
\end{equation}
We write $V = \tilde{V}$ because in this case modular parallel transport is generated entirely by scrambling modes, without non-scrambling contributions. 

To proceed, we need $D_\lambda \tilde{V}$, which was defined in equation~(\ref{dvdef}). To eludicate its meaning, let us take an ordinary $\lambda$-derivative of equation~(\ref{vexpl}): 
\begin{equation}
\frac{d\tilde{V}}{d\lambda} = 
\frac{d^2L}{d\lambda^2} \partial_L P^{\rm mod} 
+ \frac{dL}{d\lambda} \frac{d}{d\lambda} \partial_L P^{\rm mod}
- \frac{d^2R}{d\lambda^2} \partial_R P^{\rm mod}
- \frac{dR}{d\lambda} \frac{d}{d\lambda} \partial_R P^{\rm mod}
\label{dvexpl}
\end{equation}
The terms $\partial P^{\rm mod}$ are combinations of modular scrambling modes, but what about the terms $\partial^2 P^{\rm mod}$? Let us evaluate them explicitly: 
\begin{align}
\partial^2_L P^{\rm mod} & = \phantom{-}\frac{2}{R-L}\, \partial_L P^{\rm mod} 
\equiv \Gamma^L_{LL}\, \partial_L P^{\rm mod} 
\label{gammalll} \\
\partial^2_{LR} P^{\rm mod} & = -\frac{2}{(R-L)^2}\, P^{\rm mod} \\
\partial^2_R P^{\rm mod} & = - \frac{2}{R-L}\, \partial_R P^{\rm mod} 
\equiv \Gamma^R_{RR}\, \partial_R P^{\rm mod}
\label{gammarrr}
\end{align}
On the right hand side we recognize the coefficients as kinematic Christoffel symbols, which were defined in~(\ref{kschristoffels}). The term $\partial^2_{LR} P^{\rm mod}$ is not a scrambling mode but a zero mode, so it does not give rise to any non-vanishing Christoffel symbols and drops out from $D_{\lambda}\tilde{V}$. Note, however, that it also drops out\footnote{The fact that $d\tilde{V}/d\lambda$ contains no zero modes is not an accident. Whereas $\tilde{V}(\lambda)$ generates the motion from $\lambda$ to $\lambda + d\lambda$, we can think of $-\tilde{V}(\lambda-d\lambda)$ as generating the motion from $\lambda$ to $\lambda -d\lambda$. Therefore both must be free of zero modes of $H^{\rm mod}(\lambda)$, and so must $d\tilde{V}/d\lambda$.} from $d\tilde{V}/d\lambda$ so in fact $d\tilde{V}/d\lambda = D_{\lambda}\tilde{V}$.

In terms of~(\ref{gammalll}-\ref{gammarrr}), the covariant derivative of $\tilde{V}$ is:
\begin{equation}
D_\lambda \tilde{V} = 
\left(\frac{\partial^2 L}{\partial \lambda^2} 
+ \Gamma^L_{LL} \frac{\partial L}{\partial \lambda} \frac{\partial L}{\partial \lambda}\right) 
\partial_L P^{\rm mod} 
-
\left(\frac{\partial^2 R}{\partial \lambda^2} 
+ \Gamma^R_{RR} \frac{\partial R}{\partial \lambda} \frac{\partial R}{\partial \lambda}\right) 
\partial_R P^{\rm mod} 
\label{dvdlads3}
\end{equation}
We now use the ansatz $Z d\lambda = P^{\rm mod} dw$. Equation~(\ref{dzequal}) demands that the combination
\begin{equation*}
[L' \partial_L P^{\rm mod} - R' \partial_R P^{\rm mod}, P^{\rm mod} dw] 
- (L'' + \Gamma^L_{LL} L'^2) \partial_L P^{\rm mod} d\lambda
+ (R'' + \Gamma^R_{RR} R'^2) \partial_R P^{\rm mod} d\lambda
\end{equation*}
must be a multiple of (\ref{vexpl}), where primes denote $d/d\lambda$. Using the commutation relations
\begin{equation}
[\partial_L P^{\rm mod}, P^{\rm mod}] = \partial_L P^{\rm mod} 
\qquad {\rm and} \qquad 
[\partial_R P^{\rm mod}, P^{\rm mod}] = - \partial_R P^{\rm mod},
\label{comm2}
\end{equation}
which follow from~(\ref{sl2ridentification}), we obtain a linear equation for $dw/d\lambda$. Its solution reads:
\begin{equation}
Z d\lambda = \left(\frac{L''(\lambda)}{2L'(\lambda)} -\frac{R''(\lambda)}{2R'(\lambda)} 
    + \frac{R'(\lambda) + L'(\lambda)}{R(\lambda)-L(\lambda)} \right) P^{\rm mod} d\lambda
    \label{ads3result}
\end{equation}
This solution is a zero mode, like equation~(\ref{problem2}) demands. It is also unique, so it automatically satisfies demand~(\ref{problem3}). It is, in fact, the generator of bulk parallel transport along a curve $\mathcal{C}$, whose tangents span boundary intervals $(L(\lambda), R(\lambda))$. 

\begin{figure}[t]
\centering
\includegraphics[width=.66\textwidth]{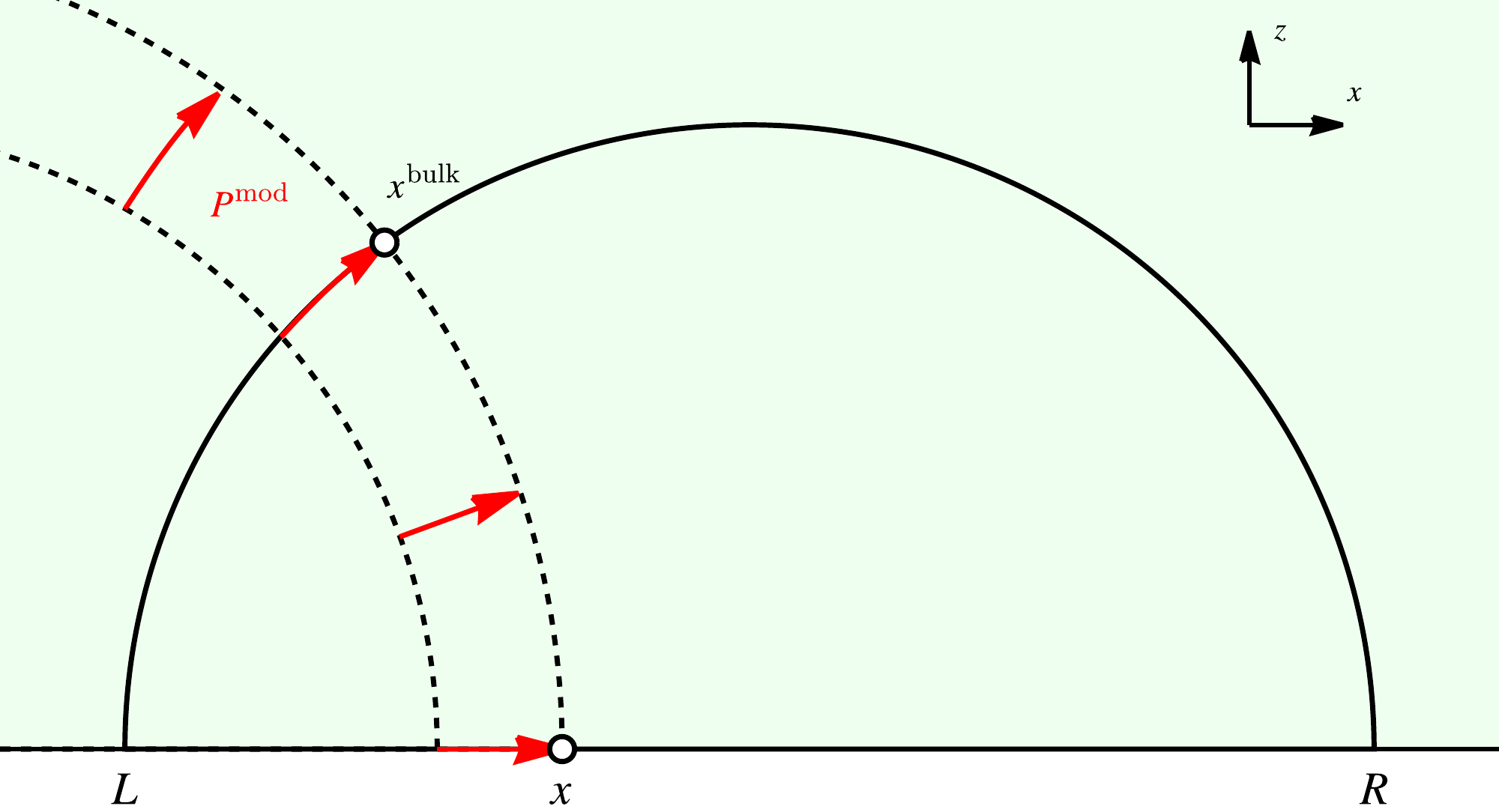}
\caption{In pure AdS$_3$, the bulk action of $P^{\rm mod}$ is to rigidly shift geodesics, which are co-planar with and orthogonal to the RT surface. Equation~(\ref{defw2}) describes the action of $P^{\rm mod} = \partial_w$ in terms of bulk intersection points of the RT surface and its orthogonal geodesics.} 
\label{fig:xbulk}
\end{figure}

\paragraph{A geometric perspective}
To understand how solution~(\ref{ads3result}) arises in a more geometric way, observe that $P^{\rm mod}$ generates bulk translations along the RT surface $a(R(\lambda))$. The bulk translation along a geodesic that subtends interval $(L,R)$ is, as a conformal symmetry of the boundary $x$-axis, a conformal translation in $w$ given by:
\begin{equation}
    e^{w}=\frac{R-x}{x-L} \quad \Longrightarrow \quad P^{\rm mod} = \partial_w
\label{defw}
\end{equation}
Note that the action of $P^{\rm mod}$ rigidly shifts geodesics, which are orthogonal to $a(R(\lambda))$; see Figure~\ref{fig:xbulk}. Therefore, we can also express $w$ using a new coordinate $x^{\rm bulk}$, which is where $a(R(\lambda))$ intersects the orthogonal geodesic anchored at $x$ on the boundary. Written in terms of $x^{\rm bulk}$, $w$ looks identical except for a factor of 2:
\begin{equation}
    e^{2w}=\frac{R-x^{\rm bulk}}{x^{\rm bulk}-L}
\label{defw2}
\end{equation}

Let $x_{\pm}^{\rm bulk}$ be the coordinates of the two bulk intersection points between consecutive geodesics along trajectory $\mathcal{C}$: $a(R(\lambda - d\lambda)) \cap a(R(\lambda))$ and $a(R(\lambda)) \cap a(R(\lambda + d\lambda))$. The multiple of $P^{\rm mod}$ that solves (\ref{dzequal}) is the distance $dw$ between $w(x_{+}^{\rm bulk})$ and $w(x_{-}^{\rm bulk})$. Using the fact that geodesics in Poincar{\'e} coordinates look like semicircles on the flat $(x,z)$-plane, we find the two intersection points by solving:
\begin{align}
    &\left(x_\pm^{\rm bulk}-L(\lambda)\right)\left(x_\pm^{\rm bulk}-R(\lambda)\right)+z^2=0 \\
    &\left(x_\pm^{\rm bulk}-L(\lambda\pm d\lambda)\right)\left(x_\pm^{\rm bulk}-R(\lambda \pm d\lambda)\right)+z^2=0
\end{align}
Expanding all terms to second order, namely
\begin{equation}
L(\lambda \pm d\lambda) = L(\lambda) \pm L'(\lambda) d\lambda + \frac{1}{2} L''(\lambda) d\lambda^2 + \ldots
\end{equation}
and likewise for $R(\lambda \pm d\lambda)$, we get:
\begin{equation}
    x_{\pm}^{\rm bulk}= \frac{L(\lambda\pm d\lambda)R(\lambda \pm d\lambda)-L(\lambda)R(\lambda)}{\left(L(\lambda\pm d\lambda)+R(\lambda \pm d\lambda)\right)-\left(L(\lambda)+R(\lambda)\right)}
\end{equation}
and consequently: 
\begin{equation}
\frac{dw}{d\lambda} = \frac{1}{2} \log{\frac{\left(R(\lambda)-x_{-}^{\rm bulk}\right) \left(x_{+}^{\rm bulk}-L(\lambda)\right)}{\left(x_{-}^{\rm bulk}-L(\lambda)\right) \left(R(\lambda)-x_{+}^{\rm bulk}\right)}} 
\end{equation}
Expanding the logarithm and ignoring terms subleading in $d\lambda$ returns equation~(\ref{ads3result}).

\paragraph{The vielbein postulate} 
We would like to relate this example to the material in Section~\ref{sec:vielbein}. There, we identified equation~(\ref{condpostulate})---a rewriting of our transport problem---as a projective generalization of the vielbein postulate~(\ref{thepostulate}). In the present example, we demanded that 
\begin{equation*}
[L' \partial_L P^{\rm mod} - R' \partial_R P^{\rm mod}, P^{\rm mod} dw] 
- (L'' + \Gamma^L_{LL} L'^2) \partial_L P^{\rm mod} d\lambda
+ (R'' + \Gamma^R_{RR} R'^2) \partial_R P^{\rm mod} d\lambda
\end{equation*}
be a multiple of $\tilde{V} = L' \partial_L P^{\rm mod} - R' \partial_R P^{\rm mod}$. From the solution~(\ref{ads3result}), we can read off that that multiple is:
\begin{equation}
-\frac{1}{2} \frac{d}{d\lambda} \log \left[\frac{1}{(R-L)^2}\frac{dR}{d\lambda}\frac{dL}{d\lambda}\right]
\label{settopostulate}
\end{equation}
Demanding the vielbein postulate instead of the weaker equation~(\ref{condpostulate}) is equivalent to setting this multiple to zero. Therefore, the equation that defines $Z d\lambda$ is equivalent to (a component of) the kinematic vielbein postulate provided that the kinematic trajectory is parameterized such that: 
\begin{equation}
d\lambda^2 = {\rm const.} \times \frac{dR\, dL}{(R-L)^2}
\label{vacuumksmetric}
\end{equation}
This is the metric of the CFT$_2$ vacuum kinematic space, which was previously discussed e.g. in \cite{intgeo, earlylampros}. 

In summary, once $\lambda$ is chosen as a length parameter in metric~(\ref{vacuumksmetric}), equation~(\ref{condpostulate}) coincides with a component of the vielbein postulate. In that circumstance, our transport problem reduces to adjusting a component of the kinematic vielbein to point in the direction of $\tilde{V}$. 

\paragraph{Heuristics: The radial cutoff line}
For a simple check of solution~(\ref{ads3result}), take the radial curve $z = z_0$. The tangent geodesics subtend $(L,R) = (\lambda - z_0, \lambda + z_0)$. We get $Z d\lambda = P^{\rm mod} d\lambda/z_0$, so the rate of translation is proportional to $1/z_0$, as expected. 

When we combine the effect of modular parallel transport with the bulk translation, we get a notion of transport that is analogous to our GPS example:
\begin{equation}
A\, d\lambda = V d\lambda + Z d\lambda = (-L_{-} + \bar{L}_-) d\lambda / z_0
\label{bulkradial}
\end{equation}
This accords with how translations are generated in the CFT, with the local length scale set by $z_0$. But in the bulk, it is neither bulk parallel transport (generated by $Z d\lambda$) nor modular parallel transport for tangent RT surfaces (generated by $V d\lambda$); instead, it is the combination of the two. We call the motion generated by $A\, d\lambda$ `sliding without rolling.' This designation captures its intuitive meaning, to be contrasted with `rolling without slipping' (modular parallel transport) and bulk parallel transport (which implicates the `Coriolis effect.') 

\subsection{Bulk parallel transport without time reflection symmetry}
\label{sec:notimerefl}

We now generalize the analysis to bulk curves $\mathcal{C}$, which do not live on a static bulk slice. This requires two new ingredients, which we discuss in turn:

\subsubsection{Tangent geodesics do not intersect. Null vector alignment}

The discussion in Section~\ref{sec:rev} considered a polygonal approximation to curve $\mathcal{C}$. The line segments that comprised the polygonal approximation were all drawn from geodesics tangent to $\mathcal{C}$. In Section~\ref{sec:mainrefl}, when we constructed a boundary operator that generates bulk parallel transport, we used this polygonal approximation in intermediate steps of the derivation. Of course, the final answer---equation~(\ref{dzequal}) and its souped-up version~(\ref{problem1}-\ref{problem3})---is independent of the polygonal approximation. 

However, when $\mathcal{C}$ does not live on a time reflection-symmetric slice of the bulk geometry, its tangent geodesics do not generically intersect. In Euclidean 3d-space, one can easily see this by inspecting the tangents to a helix $(x,y,z) = (\cos \lambda, \sin\lambda, t \lambda)$, which are:
\begin{equation}
(x,y,z) = (\cos \lambda_i, \sin\lambda_i, t \lambda_i) + s_i\, (-\sin \lambda_i, \cos\lambda_i, t)
\end{equation}
Setting equal the $x$- and $y$-coordinates of two such geodesics picks out unique values of $s_1$ and $s_2$. But then their $z$-coordinates do not match and the two tangent geodesics miss one another, separated in the `vertical' direction. The conclusion is the same when the third direction is timelike and the background is not flat, and applies to generic curves $\mathcal{C}$.

\begin{figure}[t]
\centering
\includegraphics[width=.56\textwidth]{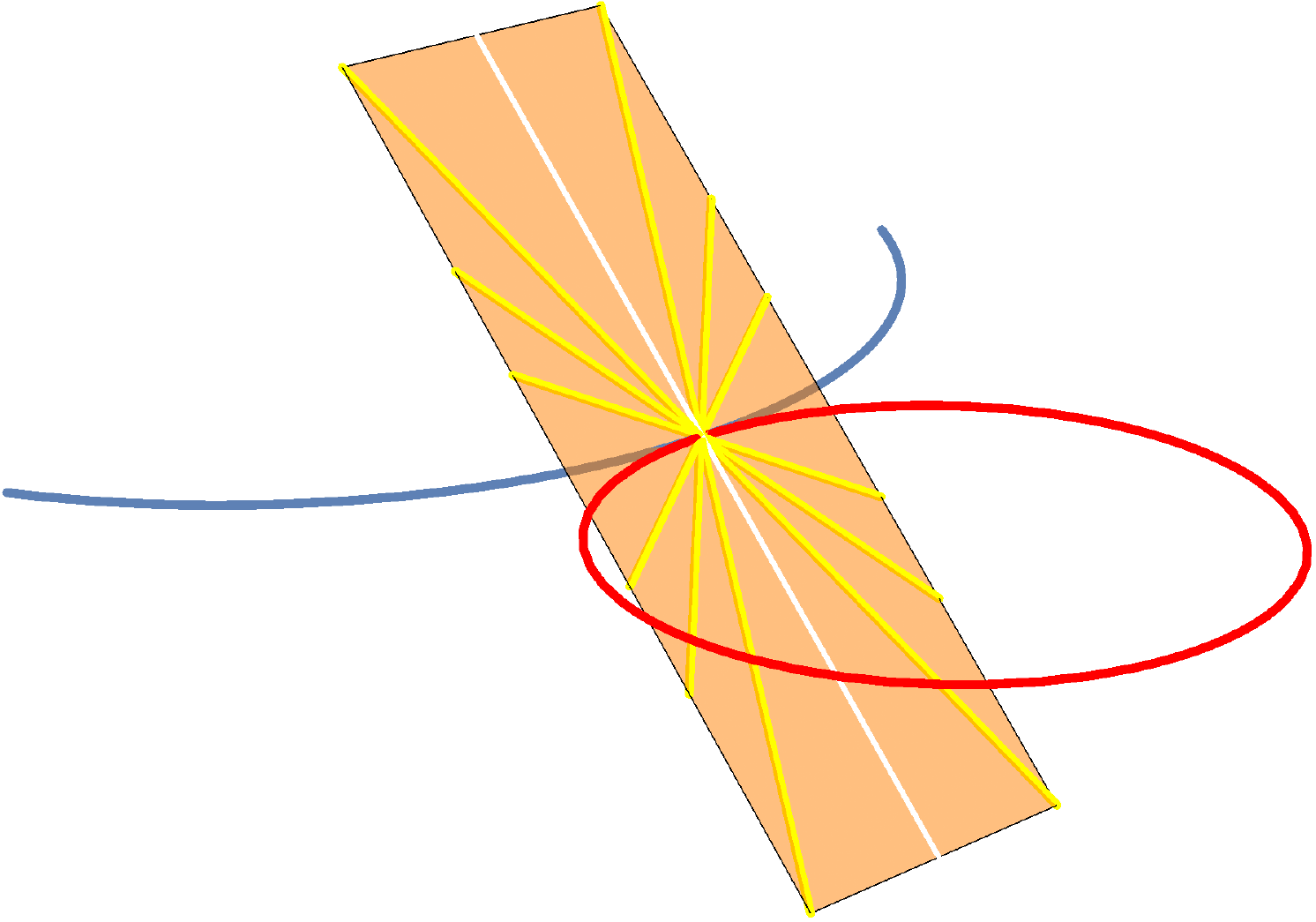}
\caption{A bulk curve (red circle) is shown with a tangent geodesic (blue) and an orthogonal lightray (white). The two locally generate a null plane (orange), which is tangent to an orthogonal lightsheet of the curve. Geodesics, which are tangent to that null plane and meet the bulk curve, are null vector-aligned (NVA); here they are shown in yellow. The figure is reproduced with permission from \cite{Czech:2019hdd}.} 
\label{fig:NVA}
\end{figure}

A consequence of this fact is that a discrete family of tangent geodesics does not give us a polygonal approximation of $\mathcal{C}$, as it did in Section~\ref{sec:rev}. This poses a technical challenge to the derivation presented in Section~\ref{sec:main}. But, in fact, the tangency condition can be relaxed in a way that restores the polygonal approximation. The requisite condition was introduced in \cite{hholes} and dubbed `null vector alignment' (NVA); see also \cite{Czech:2019hdd} for a pedagogical explanation of null vector alignment.

An NVA geodesic is (i) tangent to a lightsheet orthogonal to the curve and (ii) meets the curve $\mathcal{C}$ at one point; see Figure~\ref{fig:NVA} for an illustration of the concept. It is related to the tangent geodesic by a bulk null rotation \cite{Czech:2019hdd}, which is (locally) a symmetry of the orthogonal lightsheet. The magnitude of the null rotation is a free parameter. By adjusting this free parameter recursively, we can ensure that a sequence of geodesics which are NVA to $\mathcal{C}$ at $\lambda_1, \lambda_2, \ldots$ intersect their consecutive neighbors. If so, the line segments between consecutive intersection points do form a polygonal approximation to $\mathcal{C}$, just like we stipulated in Section~\ref{sec:rev}.

In order to extend the derivation of Section~\ref{sec:main} to arbitrary curves, we assume that we can approximate $\mathcal{C}$ to arbitrary precision by polygons in 2+1 dimensions. The line segments, which make up these polygonal approximations, are by definition NVA to $\mathcal{C}$ but not generically tangent to it. But in the continuum limit, where the polygonal approximation becomes $\mathcal{C}$ itself, these NVA geodesics will become tangent geodesics. Therefore, our final answer will only involve tangent geodesics. 

\begin{figure}[t]
\centering
\includegraphics[width=.58\textwidth]{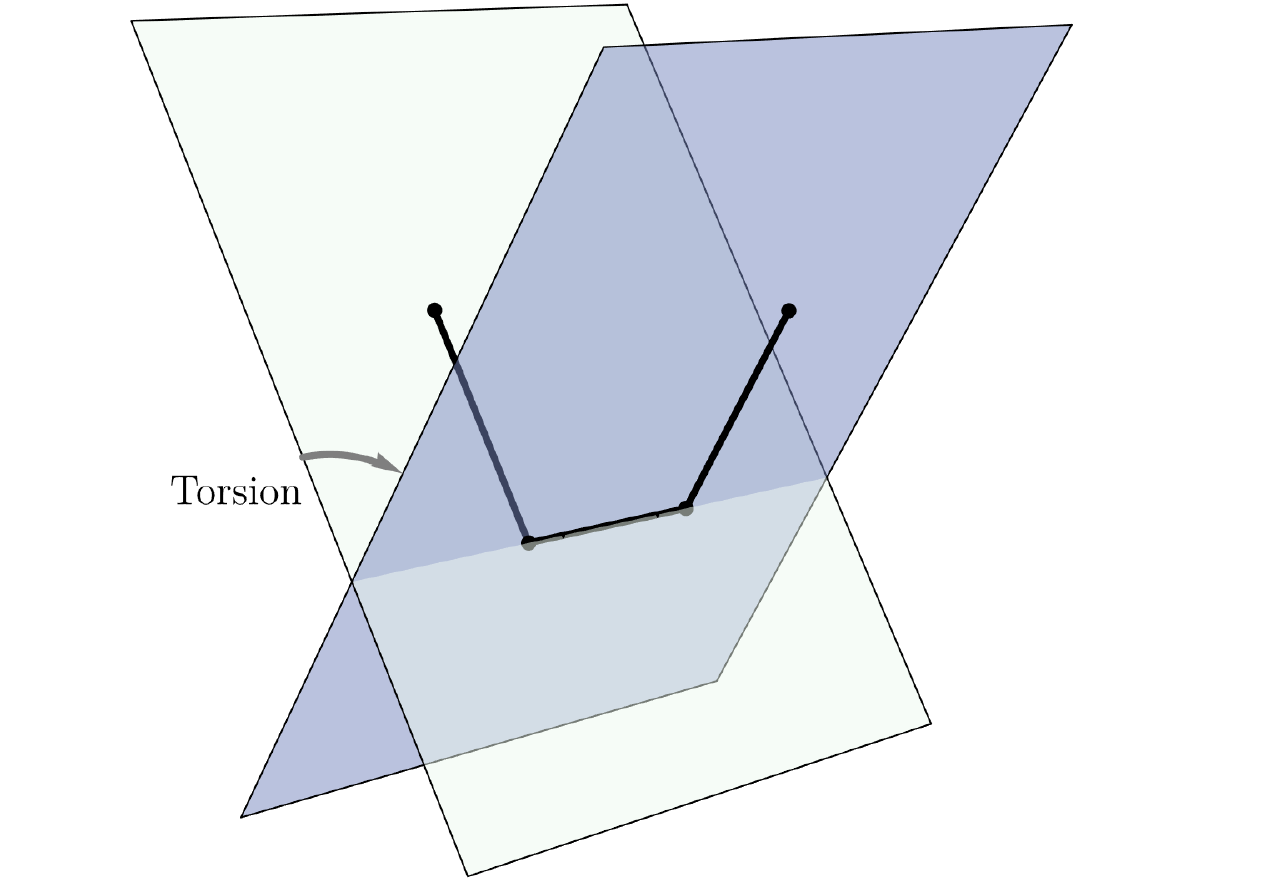}
\caption{Setup of Section~\ref{sec:torsion}: Three line segments that are part of a polygonal approximation of curve $\mathcal{C}$, which are not co-planar. Locally, the line segments define two 2-planes, which contain the segments pairwise. The angle between the two 2-planes is torsion. } 
\label{fig:torsion}
\end{figure}

\subsubsection{Torsion. How to isolate it}
\label{sec:torsion}

If $\mathcal{C}$ does not live on a time reflection-symmetric slice of the bulk geometry, its osculating plane will change. The rate of this change is called torsion. Torsion is the second complication, which we must overcome when describing bulk parallel transport along generic curves in 2+1 dimensions.

It is useful to present this complication using a polygonal approximation to $\mathcal{C}$. Let $R(\lambda)$ be a discrete family of boundary intervals whose subtending geodesics $a(R(\lambda))$ intersect in consecutive pairs. In Figure~\ref{fig:torsion} we draw three line segments, which connect consecutive intersections points of the $a(R(\lambda))$s; these line segments are part of a polygonal approximation of $\mathcal{C}$. Locally, the two-dimensional plane spanned by $a(R(\lambda - d\lambda))$ and $a(R(\lambda))$ is (approximately) an osculating plane of curve $\mathcal{C}$. The same applies to the plane spanned by $a(R(\lambda))$ and $a(R(\lambda + d\lambda))$. These two planes are generically different. The angle between them can be called $T d\lambda$, where $T$ is the `torsion' of $\mathcal{C}$. For spacelike curves in 2+1 dimensions, $T d\lambda$ is a hyperbolic angle, which represents the relative boost between the two osculating planes. 

The two osculating planes intersect on the RT surface $a(R(\lambda))$. Therefore, the hyperbolic angle between the two osculating planes ($T d\lambda$) is the magnitude of a boost about $a(R(\lambda))$. In boundary language, such boosts are generated by $H^{\rm mod}(\lambda)$. 

\paragraph{Decomposing $Z d\lambda$ into bulk translations and torsion}
In solving equations~(\ref{problem1}-\ref{problem3}), we are finding a special zero mode $Zd\lambda$, which ensures that $A d\lambda = V d\lambda + Z d\lambda$ drags the point where $a(R(\lambda))$ touches $\mathcal{C}$. We should understand this as the transport of a local coordinate system, which covers a neighborhood of $a(R(\lambda))$. (In Ref.~\cite{sewingkit} such local coordinate systems were called modular frames.) In addition to translations along $a(R(\lambda))$, however, a local coordinate system is also affected by the action of $H^{\rm mod}(\lambda)$. Unless time-reflection symmetry ensures otherwise, $Z d\lambda$ will in general contain both a translational and a torsion-like component. To find bulk translations, we must siphon out the torsional part of $Z d\lambda$. 

It is clear how to do this in pure AdS$_3$. We solve equations~(\ref{problem1}-\ref{problem3}) for $Z d\lambda$ and then write $Z d\lambda$ as a linear combination of $H^{\rm mod}(\lambda)$ and $P^{\rm mod}(\lambda)$, which was defined in equation~(\ref{defpmod}). The generator of bulk translations is the $P^{\rm mod}$-component of $Z d\lambda$. The multiple of $H^{\rm mod}$ in $Z d\lambda$ is the torsion form $T d\lambda$. 

\paragraph{Picking out the torsion component}
Away from pure AdS$_3$ (and from the CFT$_2$ ground state), how do we split $Z d\lambda$ into a translational and torsion-like component? One is tempted to say that the bulk translation is the projection of $Z d\lambda$ to the orthogonal complement of $H^{\rm mod}(\lambda)$ in the space of its zero modes. However, we do not have a preferred inner product on the commutant of $H^{\rm mod}(\lambda)$, and even if we did, the existence of orthogonal complements in infinite-dimensional vector spaces is subtle. Therefore, we would prefer a cleverer way to separate translations from torsion. 

\begin{figure}[t]
\centering
\includegraphics[width=.55\textwidth]{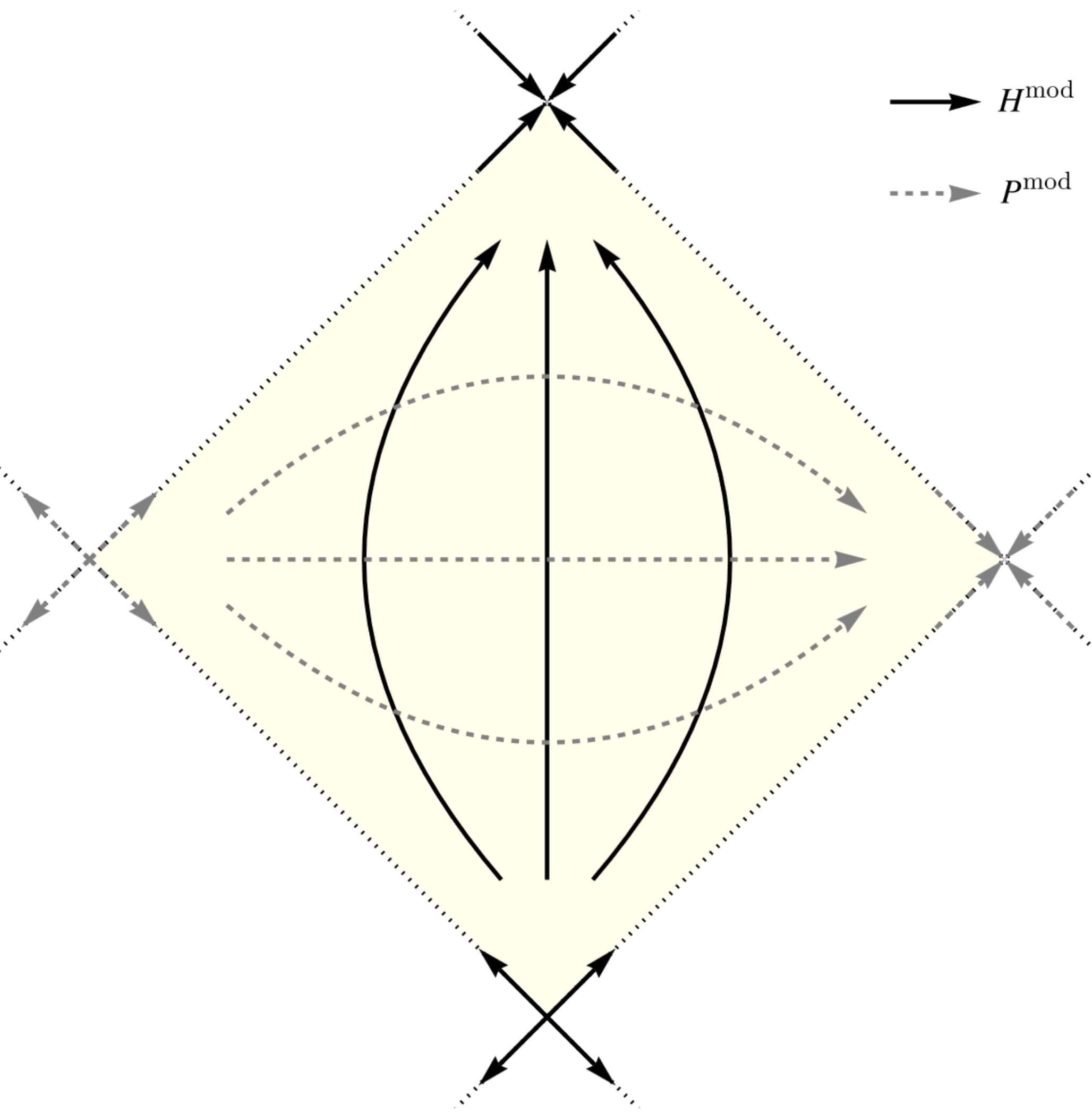}
\caption{Global symmetries $H^{\rm mod}$ and $P^{\rm mod}$ of a bulk RT surface, as seen on the boundary.} 
\label{fig:zeroes}
\end{figure}

We propose the following. Let us inspect the action of $H^{\rm mod}$ and its translational counterpart at an endpoint of interval $R(\lambda)$. In a neighborhood of an endpoint, $H^{\rm mod}$ acts as a boost whereas the translational mode acts as a dilation. We can also inspect the top and bottom of the boundary causal diamond of $R(\lambda)$; there the relation is reversed. This is summarized in Fig.~\ref{fig:zeroes}.

Consider the operator-valued form
\begin{equation}
(Z - q H^{\rm mod}) d\lambda ,
\label{defzq}
\end{equation}
where $Z d\lambda$ solves~(\ref{problem1}-\ref{problem3}). Bulk translations are generated by (\ref{defzq}) for some value of $q$. To find it, we inspect the action of (\ref{defzq}) near endpoints of $R(\lambda)$, as well as the top and bottom points of its boundary causal diamond. We demand that (\ref{defzq}) tend to a pure dilation (respectively boost) when one approaches an endpoint of $R(\lambda)$ (respectively top/bottom of its boundary causal diamond.) 

One can write this condition as an equation by considering the adjoint action of (\ref{defzq}) on a primary operator, which is inserted at one of the four corners of the boundary causal diamond of $R(\lambda)$. In particular, for scalar operators at the top or bottom of the causal diamond, we demand:
\begin{equation}
[Z - q H^{\rm mod}, \mathcal{O}({\rm top})] = 0 = [Z - q H^{\rm mod}, \mathcal{O}({\rm bottom})]
\end{equation}

\section{Discussion}
\label{sec:discussion}
In prior work on modular parallel transport, we found that holonomies of that transport are related to integral transforms of curvature invariants in the bulk \cite{sewingkit, modularchaos}. Because modular parallel transport is a boundary concept, this was an effort to relate bulk curvature to the boundary language. The present paper advances this effort in the context of the AdS$_3$/CFT$_2$ correspondence. Now, for the first time, we know what boundary operators generate bulk parallel transport along a generic spacelike curve in an asymptotically AdS$_3$ geometry. But the Riemann tensor is the commutator of coordinate vector fields. If we know how to transport along an arbitrary bulk curve, we in principle know the local Riemann tensor. 

In practice, the recovery of the bulk Riemann tensor is subject to further subtleties. For once, we have not yet discussed transport along timelike curves, which are necessary to complete this picture. As we will see momentarily, parallel transport along timelike curves comes with a unique set of challenges. Perhaps more importantly, we have not constructed the tangent space of vectors $T_x$ at a bulk point $x$. (The reader is asked not to confuse it with the tangent space to kinematic space $T_\lambda$.) We have merely looked at isolated bulk curves $\mathcal{C}$, without trying to understand the set of all curves that pass through the same bulk point.  

For a sketch of what can be accomplished with the tools in this paper, consider two intersecting spacelike curves $\mathcal{C}_1$ and $\mathcal{C}_2$ and treat them as a local coordinate grid near their intersection point $x$. In this restricted setup, the bulk Riemann tensor at the intersection point is related---in the boundary language---to the commutator of operators $Z_1$ and $Z_2$, which solve our transport problems for $\mathcal{C}_{1,2}$. To complete a calculation of the Riemann tensor, we would need to project $[Z_1, Z_2]$ to an operator subspace, which generates translations at $x$---the boundary avatar of $T_x$. Our paper does not address this last step. Doing so should be possible, for example by combining the insight of \cite{intersectinghmods} with the JLMS relation~\cite{jlmsref}. 
\smallskip

\paragraph{Assumptions}
\bartek{Our work relies on several assumptions, which we stated on the go. Here we collect them in one place for clarity:}
\begin{itemize}
\item \bartek{We of course work in holographic settings and assume large $N$.}
\item \bartek{We assume that scrambling modes (equation~\ref{scramblingmodes}) exist \cite{modularchaos}.}
\item \bartek{We assume that scrambling modes are not implicated in state deformations and only play a role in region deformations; see footnote~\ref{ftstatescr}.}
\item \bartek{We assume that projectors onto modular zero modes ($P^0[\ldots]$) and modular scrambling modes ($P^\pm[\ldots]$) exist, and ignore issues of uniqueness. (See \cite{virasoroberry} for a discussion of this point.)}
\item \bartek{Likewise, we assume that a projector onto the kernel of map $P^0_{R(\lambda)} \Big[ [\tilde{V}(\lambda), [\tilde{V}(\lambda), \,\ldots]] \Big]$ exists and ignore the issue of its uniqueness; see equation~(\ref{problem3}).}
\end{itemize}
\bartek{The last two items are important caveats. They require a detailed study, which is outside the scope of this paper. We hope to address these problems in future work.}

\bartek{Finally, we have assumed that the spacelike curve $\mathcal{C}$ under study lies not too deep in the bulk and is not too `radial'; see footnote~\ref{ftradial}. This assumption could be relaxed if we knew the analogue of modular Hamiltonians for entwinement~\cite{entwinement}---that is, boundary generators of boosts orthogonal to non-minimal geodesics. }

\bartek{A more substantial limitation of our work concerns:}

\subsection{Parallel transport along timelike curves}
This case needs special treatment because tangents to timelike curves---that is, timelike geodesics---do not reach the boundary of an asymptotically AdS spacetime and do not select regions in the boundary CFT. Consequently, a timelike curve in the bulk does not induce a trajectory in kinematic space. Extending the formalism in the present paper to timelike transport is possible but impractical and requires significant modifications. 

Our discussion of parallel transport along timelike curves starts from Reference~\cite{bulktime}, which explained the boundary origin of proper time in the bulk. A central tenet of that work is that it takes seriously the gravitational properties of massive bodies, which travel along timelike trajectories. In particular, assuming that a massive body is point-like and sees an approximately flat environment is unrealistic. A massive object cannot be confined more narrowly than its Schwarzschild radius permits, and always sources a Schwarzschild-like solution around itself. Any discussion of the worldline of a massive body necessarily implicates the Schwarzschild solution. Consequently, the most compact yet realistic object that may move along a worldline $\mathcal{C}$ is not a point-like particle but a Schwarzschild black hole.

A key innovation of Reference~\cite{bulktime} is to replace the worldline $\mathcal{C}$ of a massive particle with a tube filled with the Schwarzschild solution. Under certain technical assumptions, the authors of \cite{bulktime} found that a Schwarzschild solution could be glued continuously along the boundary of the tube with the ambient metric in which worldline $\mathcal{C}$ lives. If the Schwarzschild solution is thermodynamically stable then it is dual to the mixed state $Ue^{-H}U^\dagger$ in the boundary. Here $H$ is a dynamical Hamiltonian of the boundary theory, which generates translations in Schwarzschild time of a static black hole in pure AdS and $U$ is a unitary whose role is to prepare the kinematic state of the black hole of interest, as well as the geometric background it will propagate in. Reference~\cite{bulktime} found that the Heisenberg-picture evolution of operators near the Schwarzschild radius is generated by the modular Hamiltonian of this mixed state with a differential coefficient equal to $d(\textrm{proper time})$ along $\mathcal{C}$, assuming $U$ has certain properties. In summary, proper time along the worldline of a massive particle is the Schwarzschild time of a black hole to which that particle may collapse, which is in turn the modular time in the microscopic quantum gravity description.

\paragraph{Connection to this work} 
To an extent, the results of \cite{bulktime} solve parallel transport along worldlines of massive particles. From the GR point of view, the equations of motion of a particle characterize its backreaction on the geometry. As such, they are equivalent to finding the Schwarzschild solution stipulated in \cite{bulktime}, whose boundary dual is some state $\rho$. So long as the particle obeys its own equations of motion, bulk parallel transport is generated by $H = -\log \rho$. 

In practice, however, it is useful to consider worldlines $\mathcal{C}$, which are not geodesic.\footnote{One reason is to accommodate the probe approximation. Solving for backreaction in GR is not a practical way to find the parabola, which is traced by a projectile. Instead, we write down a differential equation, which effectively collates the parabola from a sequence of linear segments in a fixed background geometry.} Going to a piecewise-linear approximation of $\mathcal{C}$, Reference~\cite{bulktime} would give us a sequence of modular Hamiltonians $H^{\rm mod}(\lambda)$, which generate time translations along each segment. In principle, we could then solve the transport problem, which was discussed in this paper. This is how the formalism of the present paper connects with prior understanding of timelike trajectories. 

This connection is subject to three caveats:
\begin{itemize}
\item We need a new `kinematic space' of timelike geodesics. Each element in this `kinematic space' would come with a Hamiltonian $H$, which generates Schwarzschild time in a tube surrounding the geodesic. It is not clear if this space can be constructed for a general holographic bulk spacetime without recourse to bulk ingredients.

\item Supposing such $H$'s have been constructed, parallel transport along $\mathcal{C}$ will be generated by $H(\lambda)$ with some coefficient. Deploying the apparatus of this paper for the sole purpose of finding that coefficient seems like an overkill. In any event, the hard part of the problem is to find $H(\lambda)$, not the coefficient.

\item Equation~(\ref{tks}) identified the tangent space $T_\lambda$ to kinematic space with the span of modular scrambling modes. Equations~(\ref{kschristoffels}) and (\ref{dvdef}) defined the corresponding notion of covariant derivative. The motivation was to isolate variations of $H^{\rm mod}$, which can be interpreted as vector fields in the bulk. The technical fact that enabled this was the distinction between real-frequency modes of $[H^{\rm mod}, \ldots]$ and scrambling modes, whose modular frequencies are $\pm 2 \pi i$. \bartek{(See footnote~\ref{ftstatescr} for further comments on this point.)}

In contrast, the spectrum of $[H, \ldots]$ is real. For example, for the timelike geodesic at the center of pure AdS$_3$, $H$ is the global Hamiltonian of the dual CFT$_2$. A priori, it is unclear what feature distinguishes trajectory-changing modes of $[H, \ldots]$ from those, which change the state of bulk fields far away from the particle / black hole. This is because even trajectory-changing perturbations are ordinary state perturbations---unlike scrambling modes whose action is non-unitary because they shrink or enlarge the system.
\end{itemize}
The problem of extracting the `geometric' part of the modular flow in a given code subspace (choice of background) is resolved in a forthcoming publication \cite{DJJdBLL}.

\subsection{Other applications}
\label{sec:appl}
In both Sections~\ref{sec:maingen} and \ref{sec:main} the flat connection $A d\lambda = Vd\lambda + Zd\lambda$ played a more robust role than did bulk parallel transport. For example, in the generic case where the relationship between $A$ and bulk parallel transport is not simplified by a symmetry, isolating the latter from the former requires extra work. This was discussed in Section~\ref{sec:notimerefl}.

We now offer further comments on the utility of $A d\lambda$, beyond the purpose of reconstructing bulk parallel transport.

\subsubsection{Chern-Simons description of AdS$_3$ gravity}
\label{sec:csanda}
In the context of the CFT$_2$ ground state and its descendants, $A d\lambda$ is valued in the Lie algebra of $SL(2, \mathbb{R}) \times SL(2, \mathbb{R})$, which is the isometry group of AdS$_3$. As we emphasized before, $A d\lambda$ has trivial holonomies. These two features of $A d\lambda$ are reminiscent of the flat gauge fields, which describe locally AdS$_3$ spacetimes in the Chern-Simons formalism \cite{Witten:1988hc}. In fact, we may view the $A d\lambda$ studied in this paper as a generalization of on-shell Chern-Simons gauge fields, which also applies in the presence of backreacting matter fields. We explain this assertion presently.

The Chern-Simons description of AdS$_3$ gravity has two distinct gauge fields $A^{CS}$ and $\bar{A}^{\rm CS}$, each valued in the Lie algebra of $SL(2,\mathbb{R})$. On shell, they relate to the bulk dreibein $e^a_M$ and bulk spin connection $(\omega_{M})^a_b$ via:\footnote{We use uppercase Latin letters for bulk spacetime indices, to distinguish them from kinematic space indices $\mu$ and $m$.}
\begin{equation}
(A^{\rm CS}_M)^a = \omega_M^a + e_M^a / L_{\rm AdS}
\qquad {\rm and} \qquad
(\bar{A}^{\rm CS}_M)^a = \omega_M^a - e_M^a / L_{\rm AdS}\,,
\label{csfields}
\end{equation}
where $\omega_M^a \equiv (1/2) \epsilon^{abc} \eta_{cd} (\omega_{M})^d_b$. Equation~(\ref{csfields}) lists components of the $sl(2,\mathbb{R})$-valued one-form $A^{\rm CS} \equiv (A^{\rm CS}_M)^a J_a dx^M$ (respectively $\bar{A}^{\rm CS}$), where $J_a$ are the standard basis of $so(2,1) \simeq sl(2,\mathbb{R})$.

Mirroring equation~(\ref{csfields}), our construction also associates to each spacelike curve $\mathcal{C}$ not one but two flat connections---which we may call $A d\lambda$ and $\bar{A} d\lambda$. The basic reason for the appearance of two connections is that a curve $\mathcal{C}$ from $\lambda_i$ to $\lambda_f$ can also be traversed in the other direction: from $\lambda_f$ to $\lambda_i$. Let us refer to the orientation-reversed counterpart of $\mathcal{C}$ as $\bar{\mathcal{C}}$; the counterpart of $Ad\lambda = Vd\lambda + Zd\lambda$ for $\bar{\mathcal{C}}$ will be called $-\bar{A} d\lambda$. In the general case,\footnote{To see this, observe that the Wilson lines of $A$ and $\bar{A}$ are inverses of one another. If we set
\begin{equation}
{\rm Pexp} \int_{\lambda_i}^\lambda A d\lambda 
= \left( {\rm Pexp} \int_\lambda^{\lambda_i} (-\bar{A}) d\lambda\right)^{-1}
\equiv U(\lambda_i, \lambda)
\label{twotimeevols}
\end{equation}
then $A$ and $\bar{A}$ take the following forms:
\begin{equation*}
A(\lambda) = \left(\frac{d}{d\lambda} U(\lambda_i, \lambda)\right) U(\lambda_i, \lambda)^{-1}
\qquad {\rm versus} \qquad
\bar{A}(\lambda)  = 
U(\lambda, \lambda_f)^{-1} \left(\frac{d}{d\lambda} U(\lambda_i, \lambda)^{-1} \right) U(\lambda_i, \lambda_f).
\end{equation*}} $A \neq \bar{A}$.

Nevertheless, both $Ad\lambda$ and $-\bar{A}d\lambda$ have something essential in common. In both of them, their zero-mode component $Z d\lambda$ captures the translation along the bulk curve. (Here we have assumed time reflection-symmetry, i.e. ignored the torsion discussed in Section~\ref{sec:torsion}.) In this circumstance, the local generator of translations can be obtained from $Ad\lambda$ and $\bar{A}d\lambda$ via:
\begin{equation}
Z d\lambda = (A d\lambda - \bar{A} d\lambda)/2
\label{whataboutz}
\end{equation}
We can view equation~(\ref{whataboutz}) as a boundary-language, operator-valued generalization of the dreibein
\begin{equation}
e^a_M = \frac{L_{\rm AdS}}{2} (A^{\rm CS}_M - \bar{A}^{\rm CS}_M)^a ,
\end{equation}
which in the Chern-Simons formulation of gravity functions as a generator of (gauged) translations. By the same token, the scrambling mode-projected generator of modular parallel transport $\tilde{V} d\lambda$ functions much like the spin connection $\omega_M^a$. To make this identification precise, we would need three everywhere-orthogonal vector fields in the bulk geometry indexed by $a = 1,2,3$, and find $A^a$ (respectively $\bar{A}^a$) that solve our transport problem along each of their integral curves (with either orientation). This setup is possible in three dimensions because the metric can be diagonalized everywhere.

\subsubsection{Holographic states at a cutoff}
Consider a vacuum correlation function in a CFT$_2$. It can be formally rewritten in the form
\begin{align}
\langle 0 | \mathcal{O}_2 (x_2) \mathcal{O}_1 (x_1) | 0 \rangle 
& = 
\langle 0 | e^{(\infty - x_2) P}\, \mathcal{O}_1\, e^{(x_2 - x_1) P}\, \mathcal{O}_2\, e^{(x_1 - (-\infty)) P} | 0 \rangle \nonumber \\
& = {\rm Tr}\, \rho\, e^{(\infty - x_2) P}\, \mathcal{O}_1\, e^{(x_2 - x_1) P}\, \mathcal{O}_2\, e^{(x_1 - (-\infty)) P}\,,
\label{querycorr}
\end{align}
where $P = -L_{-} + \bar{L}_-$ generates spatial translations and $\rho = |0\rangle\langle 0|$. We can think of expression~(\ref{querycorr}) as a Heisenberg-picture computation, in which $P$ evolves operators in space rather than time.

Now recall that we encountered $P$ previously in equation~(\ref{bulkradial}). Up to the factor of $z_0^{-1}$, $Pd\lambda$ is the form $Ad\lambda = Vd\lambda + Zd\lambda$ that generates the `sliding without rolling' motion along the bulk curve $z = z_0$ in the bulk of AdS$_3$. This offers a compelling Heisenberg-picture interpretation of equation~(\ref{querycorr}). The equation tells us to slide along the bulk cutoff curve $z = z_0$ in search of operators, which are then merged using the OPE. The correlation function is the multiple of the identity operator, which survives this process and arrives at $x = \infty$. Note that the bulk cutoff scale $z_0$ enters this interpretation in the correct way: if we substitute $P \to A = P / z_0$ in (\ref{querycorr}), the effective distance between $x_1$ and $x_2$ will get rescaled by just the factor mandated by holographic RG. 

This suggests that we can use expressions like (\ref{querycorr}) to define correlation functions of states at an arbitrary bulk cutoff curve, including non-homogeneous cutoffs $z \neq const.$  All we need is an operator-valued one-form, which effects the `sliding without rolling' motion along the bulk cutoff curve---like $P d\lambda / z_0$ does for the cutoff curve $z = z_0$. Our paper supplies this form: it is $Ad\lambda = Vd\lambda + Zd\lambda$ for the kinematic space trajectory that is induced by the cutoff curve. In particular, the boundary regions $R(\lambda)$ visited by the trajectory have entanglement wedges, which are tangent to the bulk curve. They are---by definition---cutoff-sized regions. 

With this understanding, we write for a correlator at a cutoff:
\begin{equation}
\langle 0 | \mathcal{O}_2 (\lambda_2) \mathcal{O}_1 (\lambda_1) | 0 \rangle_{\rm cutoff} 
= {\rm Tr}\, \rho \!
\left({\rm P}\exp \int_{\lambda_2}^\infty\!\! A d\lambda\right) 
\!\mathcal{O}_2 \!
\left({\rm P}\exp \int_{\lambda_1}^{\lambda_2} \!\! A d\lambda\right) 
\!\mathcal{O}_1 \!
\left({\rm P}\exp \int_{-\infty}^{\lambda_1} \!\! A d\lambda\right) 
\label{querystate}
\end{equation}
The operators featured in the correlator are specified in terms of $\lambda$ because there is no better coordinate along an arbitrary bulk cutoff curve. More importantly, since we are working at a cutoff, the operators should not be localized more narrowly than cutoff-sized regions $R(\lambda)$. 

This method for computing vacuum correlation functions at a cutoff was previously proposed in \cite{wlnetwork}. To effect the `sliding without rolling' motion along the bulk cutoff curve, it used the $SL(2,\mathbb{R})$ Chern-Simons field $A$. Given the argument in Section~\ref{sec:csanda}, we can now extend that construction to other states with semiclassical bulk duals. Equation~(\ref{querystate}) should be valid for all correlation functions in the code subspace but not outside it. This is because our computation of $A d\lambda$ uses geometric intuition and neglects $1/N$ corrections.

\subsubsection{Metrics in kinematic space and complexity}
One motivation for considering~(\ref{querystate}) is holographic complexity. A quantum state is defined through its correlation functions. If all bulk correlation functions (using the extrapolate dictionary) can be computed by~(\ref{querystate}) then we should view it as the definition of the bulk state at a cutoff. This suggests identifying the holographic complexity of a state at a cutoff with the cost of evaluating~(\ref{querystate}). As path-ordered integrals obey
\begin{equation}
\frac{d}{d\lambda} \left({\rm P}\exp \int_{\lambda_0}^\lambda Ad\lambda \right)
= A(\lambda) \left({\rm P}\exp \int_{\lambda_0}^\lambda Ad\lambda\right),
\end{equation} 
we are effectively asking about the differential cost of applying the operator $A(\lambda)d\lambda$. 

A cost function of this type would be a monotonically increasing function along the trajectory of cutoff-sized regions $R(\lambda)$ in kinematic space. We may as well think of it as a metric in kinematic space. That is, if a reasonable metric $g^{\rm KS}$ in kinematic space could be found, we could let
\begin{equation}
\textrm{complexity(state at cutoff)}:= \oint_{R(\lambda)} \sqrt{g^{\rm KS} (\tilde{V}, \tilde{V})}\, d\lambda\,,
\label{simplecomplexity}
\end{equation}
where $\tilde{V}(\lambda) \in T_\lambda$ generates the kinematic space motion through cutoff-sized regions. At present, we do not have a proposal for such a metric. If we restrict attention to the CFT$_2$ vacuum, however, we have metric~(\ref{vacuumksmetric}), which can be motivated in many ways, including the argument involving the kinematic vielbein postulate that we presented. Substituting~(\ref{vacuumksmetric}) in (\ref{simplecomplexity}), we find \cite{wlnetwork} that the complexity of the ground state at a cutoff agrees with the Complexity = Volume conjecture of \cite{cvconj}. We plan to investigate if $g^{\rm KS}$ can be defined more robustly in future work. Relevant prior work in this direction includes \cite{modcomm}.
\medskip

\paragraph{Acknowledgements:} We have benefited from conversations with Vijay Balasubramanian, Johanna Erdmenger, Thomas Faulkner, Micha{\l} Heller, Ling-Yan (Janet) Hung, Daniel Kabat, Samuel Leutheusser, Gilad Lifschytz, Wei Song, Huajia Wang, Yixu Wang, and Anna-Lena Weigel. BCz thanks the organizers of workshop `Quantum Information and String Theory 2019' held at Yukawa Institute for Theoretical Physics, Kyoto University, where this work was initiated; the organizers of workshop `Reconstructing the Gravitational Hologram with Quantum Information' (2022) held at the Galileo Galilei Institute in Florence, where this work was partly carried out; and the University of Amsterdam for hospitality. Both BCs have been supported by the Dushi Zhuanxiang Fellowship. JdB is supported by the European Research Council under the European Union's Seventh Framework Programme (FP7/2007-2013), ERC Grant agreement ADG 834878. LL is supported by the Simons Foundation via the It from Qubit Collaboration.

\end{document}